\documentclass[11pt,psfig]{article}
\usepackage{amssymb}
\usepackage{amsmath,amsthm}
\usepackage{amsfonts}
\usepackage{graphicx}
\usepackage{graphics}
\numberwithin{equation}{section}

\begin{document}
\title{\begin{LARGE}\textbf{Quantum stabilizer codes from Abelian and non-Abelian groups association schemes
 }\end{LARGE}}
\date{20 July 2014}
\author{\textbf{A. Naghipour$^{1,2}$}\footnote{{\it Electronic addresses:} naghipour@ucna.ac.ir and a\_naghipour@tabrizu.ac.ir} \textbf{
 M. A. Jafarizadeh$^{3}$} \footnote{{\it Electronic addresses:} jafarizadeh@tabrizu.ac.ir and mjafarizadeh@yahoo.com}
 \textbf{
 S. Shahmorad$^{2}$} \footnote{{\it Electronic address:} shahmorad@tabrizu.ac.ir}\\
 [5pt]
{\it $^{1}$Department of Computer Engineering, University College of
Nabi Akram,}\\  {\it No. 1283 Rah Ahan Street,
 Tabriz, Iran}\\
  [2mm]
   {\it $^{2}$Department of Applied Mathematics,
Faculty of Mathematical Sciences, University of Tabriz,}\\
{\it 29 Bahman Boulevard, Tabriz, Iran }\\
[2mm]
 {\it $^{3}$Department of
Theoretical Physics and Astrophysics,
Faculty of Physics, University of Tabriz,}\\
{\it 29 Bahman Boulevard, Tabriz, Iran }}

%
%
%
%
%

\date{27 September 2014}
%
%
\maketitle
\bigskip
\leftskip=0pt \hrule\vskip 8pt
\begin{small}
\hspace{-.8cm}
{\bfseries Abstract}\\\\
A new method for the construction of the binary quantum stabilizer
codes is provided, where the construction is based on Abelian and
non-Abelian groups association schemes. The association schemes
based on non-Abelian groups are constructed by bases for the regular
representation from $U_{6n}$, $T_{4n}$, $V_{8n}$ and dihedral
$D_{2n}$ groups. By using Abelian group association schemes followed
by cyclic groups and non-Abelian group association schemes a list of
binary stabilizer codes up to $40$ qubits is given in tables $4$,
$5$, and $10$. Moreover, several binary stabilizer codes of
distances $5$ and $7$ with good quantum parameters is presented. The
preference of this method specially for Abelian group association
schemes is that one can construct any binary quantum stabilizer code
with any distance by using the commutative structure of association
schemes.
\\
\vspace{.3cm}
\parindent 1em
\end{small}
\vskip 10pt\hrule
\medskip
\section{\hspace*{-.5cm}Introduction}

The important class of quantum codes are stabilizer codes. The
stabilizer codes, first introduced by Gottesman [1]. These codes are
useful for building quantum fault tolerant circuits. Stabilizer code
encompasses large class of well-known quantum codes, including Shor
$9$-qubit code [6], CSS code [7], and toric code [3]. For stabilizer
codes, the error syndrome is identified by measuring the generators
of the stabilizer group. The several methods for constructing good
families of quantum codes by numerous authors over recent years have
been proposed. In [8]-[12] many binary quantum codes have been
constructed by using classical error-correcting codes, such as
Reed-Solomon codes, Reed-Muller codes, and algebraic-geometric
codes. The theory was later extended to the nonbinary case, which
authors in [13]-[15] have introduced nonbinary quantum codes for the
fault-tolerant quantum computation. Several new families of quantum
codes, such convolutional quantum codes, subsystem quantum codes
have been studied through algebraic and geometric tools and the
stabilizer method has been extended to these variations of quantum
code [16], [17].
\\
\hspace*{0.5cm} Wang et al. [21] studied the construction of
nonadditive AQCs as well as constructions of asymptotically good
AQCs derived from algebraic-geometry codes . Wang and Zhu [22]
presented the construction of optimal AQCs. Ezerman et al. [23]
presented so-called CSS-like constructions based on pairs of nested
subfield linear codes. They also employed nested codes (such as BCH
codes, circulant codes, etc.) over $\mathbb{F}_{4}$ to construct
AQCs in their earlier work [24]. The asymmetry was introduced into
topological quantum codes in [25]. Leslie [26] presented a new type
of sparse CSS quantum error-correcting code based on the homology of
hypermaps. Authors in [27] have studied the construction of AQCs
using a combination of BCH and finite geometry LDPC codes. Various
constructions of new AQCs have been studied in [28], [29]. Here in
this work the dominant underlying theme is that of constructing good
binary quantum stabilizer codes of distance $3$ and higher, e.g.,
codes with good quantum parameters based on Abelian and non-Abelian
groups association schemes. Using Abelian and non-Abelian groups
association schemes, we obtain many binary quantum stabilizer codes.
\\
\hspace*{0.5cm} An association scheme is a combinatorial object with
useful algebraic properties (see [30] for an accessible
introduction). This mathematical object has very useful algebraic
properties which enables one to employ them in algorithmic
applications such as the shifted quadratic character problem [31]. A
$d$-class symmetric association scheme ($d$ is called the diameter
of the scheme) has $d+1$ symmetric relations $R_i$ which satisfy
some particular conditions. Each non-diagonal relation $R_i$ can be
thought of as the network $(V,R_i)$, where we will refer to it as
the underlying graph of the association scheme ($V$ is the vertex
set of the association scheme which is considered as the vertex set
of the underlying graph). In [32], [33] algebraic properties of
association schemes have been employed in order to evaluate the
effective resistances in finite resistor networks, where the
relations of the corresponding schemes define the kinds of
resistances or conductances between any two nodes of networks. In
[34], a dynamical system with $d$ different couplings has been
investigated in which the relationships between the dynamical
elements (couplings) are given by the relations between the vertexes
according to the corresponding association schemes. Indeed,
according to the relations $R_i$, the so-called adjacency matrices
$A_i$ are defined which form a commutative algebra known as
Bose-Mesner (BM) algebra. Group association schemes are particular
schemes in which the vertices belong to a finite group and the
relations are defined based on the conjugacy classes of the
corresponding group. Working with these schemes is relatively easy,
since almost all of the needed information about the scheme. We will
employ the commutative structure of the association schemes in order
to the construction of binary quantum stabilizer codes, in terms of
the parameters of the corresponding association schemes such as the
valencies of the adjacency matrices $A_i$ for $i=1,...,d$. As it
will be said in Section 3, in order to construct the binary quantum
stabilizer codes, one needs a binary matrix $A=(A_1 \vert A_2)$,
such that by removing arbitrarily row or rows from $A$ one can
obtain $n-k$ independent generators. After finding the code distance
by $n-k$ independent generators one can then determine the
parameters of the associated code.
\\
 \hspace*{0.5cm} The organization of the paper is as follows. In
 section 2, we give preliminaries such as quantum stabilizer codes,
 association schemes, group association schemes, finite Abelian
 groups and finite non-Abelian groups. Section 3 is devoted to the construction of binary quantum
 stabilizer codes based on Abelian group association schemes.  In section 4,
 we construct the binary quantum stabilizer codes based on non-Abelian group association schemes.
 The paper ends with a brief conclusion.
 \\
\section{\hspace*{-.5cm} Preliminaries }
In this section, we give some preliminaries such as quantum codes
and association schemes used through the paper.
\\
\subsection{\hspace*{-.5cm} Quantum stabilizer codes}
We recall quantum stabilizer codes. For material not covered in this
subsection, as well as more detailed information about quantum error
correcting codes, please refer to [20], [36]. We employ binary
quantum error correcting codes (QECCs) defined on the complex
Hilbert space $\mathcal{H}_{2}^{\otimes n}$ where $\mathcal{H}_{2}$
is the complex Hilbert space of a single qubit $\alpha \vert 0
\rangle + \beta \vert 1 \rangle$ with $\alpha , \beta \in
\mathbb{C}$ and $ \vert \alpha \vert^{2} + \vert \beta \vert^{2}=1$.
The fundamental element of stabilizer formalism is the Pauli group
$\mathcal{G}_n$ on $n$ qubits. The Pauli group for one qubit is
defined to consist of all Pauli matrices, together with
multiplicative factors $\pm 1$, $\pm i$:
\\
\begin{equation}\label{adm1}
\mathcal{G}_1 = \{\pm I, \pm iI, \pm X, \pm iX, \pm Y, \pm iY, \pm
Z, \pm iZ\}
\end{equation}
\\
where $X , Y$ and $Z$ are the usual Pauli matrices and I is the
identity matrix. The set of matrices $\mathcal{G}_1$ forms a group
under the operation of matrix multiplication. In general, group
$\mathcal{G}_n$ consist of all tensor products of Pauli matrices on
$n$ qubits again with multiplicative factors $\pm 1$, $\pm i$.
\\
\hspace*{0.5cm} Suppose $S$ is a subgroup of $\mathcal{G}_n$ and
define $V_S$ to be the set of $n$ qubit states which are fixed by
every element of $S$. The $V_S$ is the vector \textit{space
stabilized} by $S$, and $S$ is said to be the \textit{stabilizer} of
the space $V_S$.
\\
Consider the stabilizer $ S=\langle g_1 , ... , g_l \rangle$. The
check matrix corresponding to $S$ is an $l \times 2n$ matrix whose
rows correspond to the generators $g_1$ through $g_l$; the left hand
side of the matrix contains $1$s to indicate which generators
contain $X$s, and the right hand side contains $1$s to indicate
which generators $Z$s; the presence of a $1$ on both sides indicates
a $Y$ in the generator. The $i$-th row of the check matrix is
constructed as follows: If $g_i$ contains $I$ on the $j$-th qubit
then the matrix contain 0 in $j$-th and $n+j$-th columns. If $g_i$
contains an $X$ on the $j$-th qubit then the element in $j$-th
column is 1 and in $n+j$-th column is 0. If it contains $Z$ on
$j$-th qubit then $j$-th column contains 0 and $n+j$-th element
contains 1. And in the last, if $g_i$ contains operator $Y$ on
$j$-th qubit then both $j$-th and $n+j$-th columns contain 1.
\\
The check matrix does not contain any information about overall
multiplicative factor of $g_i$. We denote by $r(g)$ a row vector
representation of operator $g$ from check matrix, which contains
$2n$ binary elements. Define $\Lambda$ as:
\\
\begin{equation}
\Lambda = \left[
            \begin{array}{cc}
              0 & I \\
              I & 0 \\
            \end{array}
          \right]_{2n\times 2n}
\hspace*{3cm}
\end{equation}
\\
where the matrices $I$ on the off-diagonals are $n \times n$.
Elements $g$ and $g'$ of the Pauli group are easily seen to commute
if and only if \hspace*{1mm}$r(g) \Lambda r(g')^{T}=0$. Therefore
the generators of stabilizer $S=\langle g_1, ... ,g_l\rangle $ with
corresponding check matrix $M$ commute if and only if $M \Lambda
M^{T}=0$. Let $S=\langle g_1, ... ,g_l\rangle $ be such that $-I$ is
not an element of $S$. Then the generators $g_i$, $i \in \{1, ...
,l\}$ are independent if and only if the rows of the corresponding
check matrix are linearly independent.\\
Suppose $C(S)$ is a stabilizer code with stabilizer $S$. We denote
by $N(S)$ a subset of $\mathcal{G}_n$, which is defined to consist
of all elements $E \in \mathcal{G}_n$ such that $EgE^{\dag}\in S$
for all $g \in S$. The following theorem specifies the correction
power of $C(S)$.
\\
\\
\textbf{Theorem 2.1.} Let $S$ be the stabilizer for a stabilizer
code $C(S)$. Suppose $\{E_j\}$ is a set of operators in
$\mathcal{G}_n$ such that $E_{j}^{\dag} E_k \notin N(S) - S$ for all
$j$ and $k$. Then $\{E_j\}$ is a correctable set of errors for the
code $C(S)$.
\\
\\
\textit{Proof}. See [36].
\\
\\
\hspace*{0.5cm} Theorem 2.1 motivates the definition of a notion of
\textit{distance} for a quantum code in analogy to the distance for
a classical code. The \textit{weight} of an error
$E\in\mathcal{G}_n$ is defined to be the number of terms in the
tensor product which are not equal to the identity. For example, the
weight of $X_1 Z_4 Y_8$ is three. The distance of stabilizer code
$C(S)$ is given by the minimum weight of an element of $N(S)-S$. In
terms of the binary vector pairs $\textbf{(a,b)}$, this is
equivalent to a minimum weight of the bitwise OR $\textbf{(a,b)}$ of
all pairs satisfying the symplectic orthogonality condition,
\\
\begin{equation}
 A_1 \textbf{b} + A_2 \textbf{a}=0,
 \hspace*{3cm}
\end{equation}
\\
which are not linear combinations of the rows of the binary check
matrix $ A= ( A_1 \vert A_2 )$.
\\
\hspace*{0.5cm} A $2$-ary quantum stabilizer code $\mathcal{Q}$,
denoted by $[[n,k,d]]_{2}$, is a $2^{k}$-dimensional subspace of the
Hilbert space $\mathcal{H}^{\otimes n}_{2}$ stabilized by an Abelian
stabilizer group $\mathcal{S}$, which does not contain the operator
$-I$ [6], and can correct all errors up to
$\lfloor\frac{d-1}{2}\rfloor$ . Explicitly
\\
\begin{equation}
 \mathcal{Q}=\{\vert \psi \rangle: s \vert \psi \rangle = \vert \psi
\rangle, \hspace*{1mm}\forall s \in \mathcal{S}\}.
 \hspace*{3cm}
\end{equation}
\\
This code, encodes $k$ logical qubits into $n$ physical qubits. The
rate of such code is $\frac{k}{n}$. Since codespace has dimension
$2^{k}$ so that we can encode $k$ qubits into it. The stabilizer
$\mathcal{S}$ has a minimal representation in terms of $n-k$
independent generators $\{ g_1, ... , g_{n-k}\hspace*{1mm} \vert
\hspace*{1mm}\forall i \in \{1, ... , n-k\}, \hspace*{1mm}g_i \in
\mathcal{S}\}$. The generators are independent in the sense that
none of them is a product of any other two (up to a global phase).
\\
\\
\hspace*{0.5cm} As in classical coding theory, there are two bounds
which have been established as necessary conditions for quantum
codes.
\\
\\
\textbf{Lemma 2.2} (quantum Hamming bound for binary case). For any
pure quantum stabilizer code $[[n,k,d]]_{2}$, we have the following
inequality
\begin{equation}
\sum_{j=0}^{[\frac{d-1}{2}]}\binom{n}{j}3^j2^k\leq2^n.
\end{equation}
\\
\textit{Proof}. See [5].
\\
\\
For any pure quantum stabilizer code with distance $3$, the quantum
Hamming bound is written by
\begin{equation}
 n-k\geq \lceil \log_{2}(3n+1)\rceil.
\end{equation}
It is also satisfied for degenerate codes of distances $3$ and $5$
[1].
\\
\\
\textbf{Lemma 2.3} (quantum Knill-Laflamme). For any quantum
stabilizer code $[[n,k,d]]_{q}$, we have
\begin{equation}
n\geq k+2d-2.
\end{equation}
\\
\textit{Proof}. See [2].
\\
\\
The class of quantum stabilizer codes is optimal in the sense that
its $k$ with fixed $n$ and $d$ is the largest.
\subsection{\hspace*{-.5cm} Association schemes}
The theory of association schemes has its origin in the design of
statistical experiments [18] and in the study of groups acting on
finite sets [35]. Besides, associations schemes are used in coding
theory [19], design theory and graph theory. One of the important
preferences of association schemes is their useful algebraic
structures that enable one to find the spectrum of the adjacency
matrices relatively easy; then, for different physical purposes, one
can define particular spin Hamiltonians which can be written in
terms of the adjacency matrices of an association scheme so that the
corresponding spectra can be determined easily. The reader is
referred to [4] for further information on association schemes.
\\
\\
\textbf{Definition \hspace*{1mm}2.2.1.} A d-class association scheme
$\Omega$ on a finite set $V$ is an order set $\{R_0,R_1, ... ,R_d\}$
of relations on the set $V$ which satisfies the following axioms:
\\
\\
(1)\hspace*{1mm}$\{R_0,R_1, ... ,R_d\}$ is a partition of $V\times
V$.
\\
\\
(2) $R_0$ is the identity relation, i.e., $(x,y)\in R_0$ if and only
if $x=y$, whenever $x,y \in V$.
\\
\\
(3) Every relation $R_i$ is symmetric, i.e., if $(x,y) \in R_i$ then
also $(y,x) \in R_i$, for every $x,y \in V$.
\\
\\
(4) Let $0\leq i,j,l \leq d$. Let $x,y \in V$ such that $(x,y) \in
R_l$, then the number
\\
$$ p_{ij}^{l}= \vert \{z \in V : (x,z) \in R_i \hspace*{1mm}\textrm{and}\hspace*{1mm} (z,y) \in R_j
\}\vert$$
\\
only depends on $i,j$ and $l$.
\\
\\
The relations $R_0,R_1, ... ,R_d$ are called the associate classes
of the scheme; two elements $x,y \in V$ are $i$-th associates if
$(x,y) \in R_i$. The numbers $p^{l}_{ij}$ are called the
\textit{intersection numbers} of $\Omega$. If
\\
\\
$(3)^{'} \hspace*{3mm}   R_i^{t}=R_i \hspace*{3mm}\textrm{for}
\hspace*{3mm}0\leq i\leq d,\hspace*{3mm} \textrm{where}\hspace*{3mm}
R_{i}^{t}=\{(\beta,\alpha): (\alpha,\beta) \in R_i\}$
\\
\\
then the corresponding association scheme is called symmetric.
Further, if $p^{l}_{ij}=p^{l}_{ji}$ for all $ 0\leq i,j,l \leq d$,
then $\Omega =(V,\{R_i\}_{0\leq i \leq d})$ is called commutative.
Let $\Omega$ be a commutative symmetric association scheme of class
$d$; then the matrices $A_0,A_1,...,A_d$ defined by
\\
\begin{equation}
(A_{i})_{\alpha , \beta}= \left\{
                            \begin{array}{ll}
                              1 & \hbox{if}\hspace*{2mm}(\alpha , \beta) \in R_i, \\

                              0 & \hbox{otherwise}
                            \end{array}
                          \right.
\hspace*{4cm}
\end{equation}
\\
are adjacency matrices of $\Omega$ and are such that
\\
\begin{equation}
A_i A_j = \sum_{l=0}^{d}p_{ij}^{l}A_l. \hspace*{4cm}
\end{equation}
\\
From (2.9), it is seen that the adjacency matrices $A_0,A_1, ...
,A_d$ form a basis for a commutative algebra $\mathbf{A}$ known as
the Bose-Mesner algebra of $\Omega$. This algebra has a second basis
$E_0, ... E_d$ primitive idempotents,
\\
\begin{equation}
E_0 =\frac{1}{n}J,\hspace*{3mm}E_i E_j=\delta_{ij}
E_i,\hspace*{3mm}\sum_{i=0}^{d}E_i = I,
 \hspace*{3cm}
\end{equation}
\\
where $\nu =\vert V\vert$ and $J$ is an $\nu \times \nu$ all-one
matrix in $A$. In terms of the adjacency matrices $A_0,A_1, ...
,A_d$ the four defining axioms of a $d$-class association scheme
translate to the following four statements [39]:
\\
\begin{equation}
\sum_{l=0}^{d}A_l=J,\hspace*{3mm}A_0=I,\hspace*{3mm}A_i =A_i^{T}
\hspace*{3mm}\textrm{and} \hspace*{3mm}A_i A_j
=\sum_{l=0}^{d}p_{ij}^{l}A_l.
 \hspace*{3cm}
\end{equation}
\\
\\
with $0\leq i,j \leq d$ and where $I$ denotes the $\nu \times \nu$
identity matrix and $A^{T}$ is the transpose of $A$. Consider the
cycle graph with $\nu$ vertices by $C_\nu$. It can be easily seen
that, for even number of vertices $\nu=2m$, the adjacency matrices
are given by
\\
\begin{equation}
A_i = S^{i}+ S^{-i},\hspace*{3mm} i=1,2, ... ,m-1,
\hspace*{3mm}A_m=S^{m},
 \hspace*{3cm}
\end{equation}
\\
where $S$ is an $\nu \times \nu$ circulant matrix with period $\nu
(S^{\nu}= I_\nu)$ defined as follows:
\\
\begin{equation}
S=\left(
    \begin{array}{ccccccc}
      0 & 0 & 0 & \ldots & 0 & 0 & 1 \\
      1 & 0 & 0 & \ldots & 0 & 0 & 0 \\
      0 & 1 & 0 & \ldots & 0 & 0 & 0 \\
      \vdots\\
      0 & 0 & 0 & \ldots & 1 & 0 & 0 \\
      0 & 0 & 0 & \ldots & 0 & 1 & 0 \\
    \end{array}
  \right).
\hspace*{3cm}
\end{equation}
\\
For odd number of vertices $\nu =2m+1$, we have
\\
\begin{equation}
A_i = S^{i}+ S^{-i},\hspace*{3mm} i=1,2, ... ,m-1,m. \hspace*{3cm}
\end{equation}
\\
One can easily check that the adjacency matrices in (2.12) together
with $A_0=I_{2m}$ (and also the adjacency matrices in (2.14)
together with $A_0=I_{2m+1}$) form a commutative algebra.
\\
\subsection{\hspace*{-.3cm}Group association schemes}
In order to construct quantum stabilizer codes, we need to study the
group association schemes. Group association schemes are particular
association schemes for which the vertex set contains elements of a
finite group $G$ and the relations $R_i$ are defined by
\begin{equation}
R_i=\{(\alpha,\beta):\alpha \beta^{-1} \in C_i\},
 \hspace*{3cm}
\end{equation}
\\
where $C_{0}=\{e\},C_{1},\ldots,C_{d} $ are the set of conjugacy
classes of $G$. Then, $\Omega = (G,\{R_i\}_{0\leq i \leq d})$
becomes a commutative association scheme and it is called the group
association scheme of the finite group $G$. It is easy to show that
the $\textit{i}$\hspace*{0.3mm}th adjacency matrix is a summation
over elements of the $\textit{i}$\hspace*{0.3mm}th stratum group. In
fact by the action of $\bar{C}_{i}:=\Sigma_{g\in C_i}{g}$
($\bar{C}_i$ is called the $\textit{i}$\hspace*{0.3mm}th
$\textit{class sum}$) on group elements in the regular
representation we observe that $\forall \alpha , \beta ,
(\bar{C}_i)_{\alpha\beta}= (A_i)_{\alpha\beta}$, so
\\
\begin{equation}
A_i =\bar{C}_i = \sum_{g \in C_i}g,
 \hspace*{3cm}
\end{equation}
\\
Thus due to (2.9),
\begin{equation}
\bar{C}_i \bar{C}_j= \sum^{d}_{l=0}p^{l}_{ij} \bar{C}_l,
 \hspace*{3cm}
\end{equation}
\\
However the intersection numbers $p^{l}_{ij}, i,j,l =0,1,...,d$ are
given by [38]
\\
\begin{equation}
p^{l}_{ij}= \frac{\vert C_i \vert \vert C_j \vert}{\vert G \vert}
\sum^{d}_{m=0} \frac{\chi_m (g_i)\chi_m
(g_j)\overline{\chi_m(g_l)}}{\chi_m (1)},
 \hspace*{3cm}
\end{equation}
where $n:= \vert G \vert $ is the total number of vertices.
\\
\subsection{\hspace*{-.3cm}Finite Abelian groups}
The classification of finite groups is extremely difficult, but the
classification of finite Abelian is not so difficult. It turns out
that a fine Abelian group is isomorphic to a product of cyclic
groups, and there's a certain uniqueness to this representation.
\\
\subsubsection{\hspace*{-.3cm}Cyclic groups and subgroups}
Let $G$ be a group and $a\in G$. The subset
\begin{equation}
\langle a \rangle = \{a^{n}\vert n \in \mathbb{Z}\}
\end{equation}
\\
is a subgroup of $G$. It is called a $\textit{cyclic subgroup}$ of
$G$, or the subgroup $\textit{generated}$ by $a$. If $G=\langle a
\rangle$ for some $a\in G$ then we call $G$ a cyclic group.
\\
\\
The \textit{order} of an element $a$ in a group is the least
positive integer $n$ such that $a^{n}=1$. It's denoted ord $a$.
\\
\\
We will often denote the abstract cyclic group of order $n$ by $C_n
= \{1,a,a^{2}, ... , a^{n-1}\}$ when the operation is written
multiplicatively. It is isomorphic to the underlying additive group
of the ring $\mathbb{Z}_n$ where an isomorphism is
$f:\mathbb{\mathbb{Z}}_n \rightarrow C_n$ is defined by
$f(k)=a^{k}$.
\\
\\
$\hspace*{3mm}$Note that cyclic group are all Abelian, since $a^{n}
a^{m}=a^{m+n}= a^{m}a^{n}$. The integers $\mathbb{Z}$ under addition
is an infinite cyclic group, while $\mathbb{Z}_n$, the integers
modulo $n$, is a finite cyclic group of order $n$. Every cyclic
group is isomorphic either to $\mathbb{Z}$ or to $\mathbb{Z}_n$ for
some $n$.
\\
\subsubsection{\hspace*{-.3cm}Product of groups}
Using multiplicative notation, if $G$ and $H$ are two groups then $G
\times H$ is a group where the product operation $(a,b)(c,d)$ is
defined by $(ac,bd)$, for all $a,c\in G$ and all $b,d\in H$.
\\
\\
The product of two Abelian groups is also called their
\textit{direct sum}, denoted $G \oplus H$. Since every cyclic group
of order $n$ is given by the modular integers $\mathbb{Z}_n$ under
addition mod $n$. Hence, to illustrate, an Abelian group of order
1200 may actually be isomorphic to, say, the group
$\mathbb{Z}_{40}\times \mathbb{Z}_{6}\times \mathbb{Z}_{5}$.
Furthermore, the Chinese remainder theorem, as we'll see, which says
that if $m$ and $n$ are relatively prime, then $\mathbb{Z}_{_{mn}}
\cong \mathbb{Z}_{_{m}} \times \mathbb{Z}_{_{m}}$. In the preceding
example, we may then replace $\mathbb{Z}_{_{40}}$ by
$\mathbb{Z}_{_{2^{3}}}\times \mathbb{Z}_{5}$, and $\mathbb{Z}_{6}$
by $\mathbb{Z}_{2} \times \mathbb{Z}_{3}$. Therefore, we will state
the fundamental theorem like this: every finite Abelian group is the
product of cyclic groups of prime power orders. The collection of
these cyclic groups will be determined uniquely by the group $G$.
\\
\\
\textbf{Theorem 2.4} (Chinese remainder theorem for groups). Suppose
that $n=km$ where $k$ and $m$ are relatively prime. Then the cyclic
group $C_n$ is isomorphic to $C_k \times C_m$. More generally, if
$n$ is the product $k_1 \cdots k_r$ where the factors are pairwise
relatively prime, then
\begin{equation}
C_n \cong C_{k_{1}} \times ... \times C_{k_{r}}=
\prod^{r}_{i=1}C_{k_{_{i}}}.
 \hspace*{3cm}
\end{equation}
\\
In particular, if $n= p^{e_{1}}_{1} ... p^{e_{r}}_{r}$, then the
cyclic group $C_n$ factors as the product of the cyclic groups
$C_{p_{i}^{e_{i}}}$, that is
\\
\begin{equation}
C_n \cong \prod^{r}_{i=1} C_{p_{i}^{e_{i}}}.
 \hspace*{3cm}
\end{equation}
\\
\\
\textbf{Theorem 2.5} (Fundamental theorem of finite Abelian groups).
Every finite Abelian group is isomorphic to the direct product of a
unique collection of cyclic groups, each having a prime power order.
\\
\\
\textit{Proof}. See [37].
\\
\\
\hspace*{3mm}For the determination of the number of distinct Abelian
group of order $n$ we need to study the partition function. In
number theory, a partition of a positive integer $n$ is a way of
writing $n$ as a sum of position integers. The number of different
partitions of $n$ is given by the partition function $p(n)$ [40].
\\
\\
For instance $p(5)=7$, having seen the seven ways we can partition
5, i.e.,
\\
\\
\begin{equation}
\begin{array}{c}
5=5 \\
\hspace*{7mm}5=4+1 \\
\hspace*{7mm}5=3+2 \\
\hspace*{14mm}5=3+1+1 \\
\hspace*{14mm}5=2+2+1 \\
\hspace*{21mm}5=2+1+1+1\\
\hspace*{28mm}5=1+1+1+1+1
\end{array}
 \hspace*{3cm}
\end{equation}
\\
\\
So, there are seven Abelian group of order 32, i.e.,
\\
\\
\begin{equation}
\begin{array}{c}
  G_1=\mathbb{Z}_{2^{5}}\\
 \hspace*{9mm} G_2=\mathbb{Z}_{2^{4}}\times \mathbb{Z}_{2}\\
 \hspace*{10.5mm} G_3=\mathbb{Z}_{2^{3}}\times \mathbb{Z}_{2^{2}} \\
  \hspace*{17.7mm}G_4=\mathbb{Z}_{2^{3}}\times \mathbb{Z}_{2}\times \mathbb{Z}_{2} \\
  \hspace*{19.2mm}G_5=\mathbb{Z}_{2^{2}}\times \mathbb{Z}_{2^{2}}\times \mathbb{Z}_{2}  \\
  \hspace*{26.7mm}G_6=\mathbb{Z}_{2^{2}}\times \mathbb{Z}_{2}\times \mathbb{Z}_{2}\times \mathbb{Z}_{2} \\
  \hspace*{34mm}G_7=\mathbb{Z}_{2}\times \mathbb{Z}_{2}\times \mathbb{Z}_{2}\times
\mathbb{Z}_{2}\times \mathbb{Z}_{2}
\end{array}
\hspace*{2.2cm}
\end{equation}
\\
\\
The above function enables us to better express the number of
distinct Abelian group of a given order, as follows.
\\
\\
\textbf{Theorem 2.6.} Let $n$ denote a positive integer which
factors into distinct prime powers, written $n=\prod p_{k}^{e_{k}}$.
Then there are exactly $\prod p(e_k)$ distinct Abelian group of
order $n$.
\\
\\
\hspace*{3mm}In particular, when $n$ is square-free, i.e., all
$e_k=1$ then there is a unique Abelian group of order $n$ given by
$\mathbb{Z}_{p_{1}} \times \mathbb{Z}_{p_{2}}\times ... \times
\mathbb{Z}_{p_{k}}$, which is just the cyclic group
$\mathbb{Z}_{n}$, if we may borrow Chinese reminder theorem again.
\\
\\
\subsection{\hspace*{-.3cm}Finite non-Abelian groups}
A non-Abelian group, also sometimes called a non-commutative group,
is a group $(G,*)$ in which there are at least two elements $a$ and
$b$ of $G$ such that $a*b\neq b*a$. Non-Abelian groups are pervasive
in mathematics and physics. Both discrete groups and continuous
groups may be non-Abelian. Most of the interesting Lie groups are
non-Abelian, and these play an important role in gauge theory.
\subsubsection{\hspace*{-.3cm}Non-Abelian group $U_{6n}$}
The group $U_{6n}$, where $n\geq 1$, is generated by two generators
$a$ and $b$ with the following relations:
\begin{equation}
U_{6n}=\{a,b: a^{2n}=b^3=1, a^{-1}ba=b^{-1}\}.
 \hspace*{3cm}
\end{equation}
The group $U_{6n}$ has $3n$ conjugacy class. The $3n$ conjugacy
classes are given by, for $0\leq r\leq n-1$,
\begin{equation}
\{a^{2r}\}, \{ba^{2r}, b^2a^{2r}\}, \{a^{2r+1}, ba^{2r+1},
b^2a^{2r+1}\}.
 \hspace*{3cm}
\end{equation}
The number of group elements of $U_{6n}$ is $6n$ and the matrix
representations of $[a]$ and $[b]$ with respect to the basis
$\mathcal{B}=\{a^j, ba^j, b^2a^j\}$, for $0\leq j\leq 2n-1$, are
given by
\begin{equation}
[a]=\left(
    \begin{array}{ccc}
      S & 0 & 0 \\
      0 & 0 & S \\
      0 & S & 0 \\
     \end{array}
  \right), [b]=\left(
    \begin{array}{ccc}
      0 & I & 0 \\
      0 & 0 & I \\
      I & 0 & 0 \\
     \end{array}
  \right)
\end{equation}
where $I$ is an $2n \times 2n$ identity matrix and $S$ is an $2n
\times 2n$ circulant matrix with period $2n (S^{2n}= I_{2n})$. The
adjacency matrices $A_0$,$A_1$,...,$A_{3n-1}$ of this group are
given by
\begin{align}
A_r &=[a]^{2r}, \qquad r=0,1,...,n-1 \nonumber \\
A_{n+r} &=[b][a]^{2r}+[b]^2[a]^{2r}, \qquad r=0,1,...,n-1 \\
A_{2n+r} &=[a]^{2r+1}+[b][a]^{2r+1}+[b]^2[a]^{2r+1}, \qquad
r=0,1,...,n-1 \nonumber.
 \hspace*{3cm}
\end{align}
One can easily that the adjacency matrices in (2.27) form a
commutative algebra [4].
\subsubsection{\hspace*{-.3cm}Non-Abelian group $T_{4n}$}
The group $T_{4n}$, where $n\geq 1$, with two generators $a$ and
$b$, obeys the following relations:
\begin{equation}
T_{4n}=\{a,b: a^{2n}=1, a^{n}=b^2, b^{-1}ab=a^{-1}\}.
 \hspace*{3cm}
\end{equation}
The group $T_{4n}$ has $n+3$ conjugacy class. The $n+3$ conjugacy
classes are given by
\begin{equation}
\{1\}, \{a^{n}\}, \{a^r, a^{-r}\}(1\leq r\leq n-1), \{ba^{2j}: 0\leq
j\leq n-1\}, \{ba^{2j+1}: 0\leq j\leq n-1\}.
 \hspace*{3cm}
\end{equation}
The number of group elements of $T_{4n}$ is $4n$ and the matrix
representations of $[a]$ and $[b]$ with respect to the basis
$\mathcal{B}=\{a^j, ba^j, b^2a^j, b^3a^j\}$, for $0\leq j\leq n-1$,
are given by
\begin{equation}
[a]=\left(
    \begin{array}{cc}
      S & 0 \\
      0 & S^{-1} \\
      \end{array}
  \right), [b]=\left(
    \begin{array}{cccc}
      0 & 0 & I & 0 \\
      0 & 0 & 0 & I \\
      0 & I & 0 & 0 \\
      I & 0 & 0 & 0 \\
     \end{array}
  \right)
\end{equation}
where $I$ is an $n \times n$ identity matrix and $S$ is an $2n
\times 2n$ circulant matrix with period $2n (S^{2n}= I_{2n})$. The
adjacency matrices $A_0$,$A_1$,...,$A_{n+2}$ of this group are given
by
\begin{align}
A_{0} &=I_{4n}, \qquad n=2,3,... \nonumber \\
A_{1} &=[a]^n, \qquad n=2,3,... \nonumber \\
A_{j+1} &=[a]^{j}+[b]^2[a]^{n-j}, \qquad j=1,...,n-1 \nonumber \\
A_{n+1} &=\sum_{j=0}^{\lceil
\frac{n}{2}\rceil-1}([b][a]^{2j}+[b]^3[a]^{2j}), \qquad 2j<n  \\
A_{n+2} &=\sum_{j=0}^{\lceil
\frac{n-1}{2}\rceil-1}([b][a]^{2j+1}+[b]^3[a]^{2j+1}), \qquad 2j+1<n.
\nonumber
 \hspace*{3cm}
\end{align}
One can easily prove that the adjacency matrices in (2.31) form a
commutative algebra [4].
\subsubsection{\hspace*{-.3cm}Non-Abelian group $V_{8n}$}
The group $V_{8n}$, where $n$ is an odd integer number [38], is
generated by two generators $a$ and $b$ with the following
relations:
\begin{equation}
V_{8n}=\{a,b: a^{2n}=b^4=1, ba=a^{-1}b^{-1}, b^{-1}a=a^{-1}b\}.
 \hspace*{3cm}
\end{equation}
The group $V_{8n}$ has $2n+3$ conjugacy class. The $2n+3$ conjugacy
classes are given by
\begin{align}
\{1\}, \{b^2\}, \{a^{2r+1}, b^2a^{-2r-1}\}(0\leq r\leq
n-1),\nonumber \\
\{a^{2s}, a^{-2s}\}, \{b^2a^{2s}, b^2a^{-2s}\}(1\leq s\leq
\frac{n-1}{2}), \\
\{b^ka^j: j\; \text{even} ,\; k=1,3\}, \; \text{and} \; \{b^ka^j:
j\; \text{odd}, \;k=1,3\}\nonumber.
\end{align}
The number of group elements of $V_{8n}$ is $8n$ and the matrix
representations of $[a]$ and $[b]$ with respect to the basis
$\mathcal{B}=\{a^j, ba^j, b^2a^j, b^3a^j\}$, for $0\leq j\leq 2n-1$,
are given by
\begin{equation}
[a]=\left(
    \begin{array}{cccc}
      S & 0 & 0 & 0 \\
      0 & 0 & 0 & S^{-1} \\
      0 & 0 & S & 0 \\
      0 & S^{-1} & 0 & 0 \\
     \end{array}
  \right), [b]=\left(
    \begin{array}{cccc}
      0 & I & 0 & 0 \\
      0 & 0 & I & 0 \\
      0 & 0 & 0 & I \\
      I & 0 & 0 & 0 \\
     \end{array}
  \right)
\end{equation}
where $I$ is an $2n \times 2n$ identity matrix and $S$ is an $2n
\times 2n$ circulant matrix with period $2n (S^{2n}= I_{2n})$. The
adjacency matrices $A_0$,$A_1$,...,$A_{2n+3}$ of this group are
given by
\begin{align}
A_0 &=I_{8n}, \nonumber \\
A_{1} &=[b]^2, \nonumber \\
A_{2+j} &=[a]^{2j+1}+[b]^2[a]^{2n-2j-1}, \qquad j=0,1,...,n-1 \nonumber\\
A_{n+1+j} &=[a]^{2j}+[a]^{2n-2j}, \qquad j=1,2,...,\frac{n-1}{2}
\nonumber\\
A_{n+1+\frac{n-1}{2}+j} &=[b]^2[a]^{2j}+[b]^2[a]^{2n-2j}, \qquad
j=1,2,...,\frac{n-1}{2}\\
A_{2n+1} &=\sum_{j=0}^{n-1}([b][a]^{2j}+[b]^3[a]^{2j}), \nonumber\\
A_{2n+2} &=\sum_{j=0}^{n-1}([b][a]^{2j+1}+[b]^3[a]^{2j+1}).
\nonumber
 \hspace*{3cm}
\end{align}
One can easily prove that the adjacency matrices in (2.35) form a
commutative algebra [4].
\subsubsection{\hspace*{-.3cm}The dihedral group $D_{2n}$}
The dihedral group $G=D_{2n}$ is generated by two generators $a$ and
$b$ with the following relations:
\begin{equation}
D_{2n}=\{a,b: a^{n}=b^2=1, b^{-1}ab=a^{-1}\}
 \hspace*{3cm}
\end{equation}
We consider the case of $n=2m$; the case of odd $n$ can be
considered similarly. The dihedral group $G=D_{2n}$ with $n=2m$ has
$m+3$ conjugacy classes, are given by
\begin{align}
\{1\}, \{a^{r}, a^{-r}\}(1\leq r\leq
m-1),\nonumber \\
\{a^{m}\}, \{a^{2j}b\}(0\leq j\leq m-1), \\
\{a^{2j+1}b\}(0\leq j\leq m-1).\nonumber
\end{align}
The adjacency matrices $A_0$,$A_1$,...,$A_{m+2}$ of this group with
$n=2m$ are given by
\begin{align}
A_{0} &=I_{2n}, \nonumber \\
A_{j} &=I_{2}\otimes (S^{j}+S^{-j}), \qquad j=1,2,...,m-1 \nonumber \\
A_{m} &=I_{2}\otimes S^{m}, \nonumber \\
A_{m+1} &=\sigma_{x}\otimes(\sum_{j=0}^{m-1}S^{2j}), \\
A_{m+2} &=\sigma_{x}\otimes(\sum_{j=0}^{m-1}S^{2j+1}). \nonumber
 \hspace*{3cm}
\end{align}
where $S$ is an $n \times n$ circulant matrix with period $n (S^{n}=
I_{n})$ and $\sigma_{x}$ is the Pauli matrix. Also, the adjacency
matrices of this group with $n=2m+1$ are given by
\begin{align}
A_{0} &=I_{2n}, \nonumber \\
A_{j} &=I_{2}\otimes (S^{j}+(S^{-1})^j), \qquad j=1,2,...,m \\
A_{m+1} &=\sigma_{x}\otimes J_n. \nonumber
 \hspace*{3cm}
\end{align}
where $S$ is an $n \times n$ circulant matrix with period $n (S^{n}=
I_{n})$ and $J_{n}$ is the $n \times n$ all-one matrix. One can
easily prove that the adjacency matrices in (2.38) and (2.39) form a
commutative algebra [4].
\section{\hspace*{-.3cm}Construction of stabilizer codes from Abelian group association schemes}
To construct a quantum stabilizer code of length $n$ based on the
Abelian group association schemes we need a binary matrix $A=(A_1
\vert A_2)$ which has $2n$ columns and two sets of rows, making up
two $n \times n$ binary matrices $A_1$ and $A_2$, such that by
removing arbitrarily row or rows from $A$ we can achieve $n-k$
independent generators. After finding the code distance by $n-k$
independent generators we can then determine the parameters of the
associated code. The parameters $[[ n,k,d ]]_{2}$ of the associated
quantum stabilizer are its length $n$, its dimension $k$, and its
minimum distance $d$.
\\
\\
Consider the cycle graph $C_\nu$ with $\nu$ vertices, as is
presented in section 2.2. By setting $m=2$ in view of (2.14), we
have
\\
\begin{equation}
A_0 = I_5,\hspace*{4mm} A_1 = S+S^{-1},\hspace*{4mm} A_2=S^{2}
+S^{-2} \hspace*{3cm}
\end{equation}
\\
where $S$ is an $5\times5$ circulant matrix with period 5($S^5=I_5$)
defined as follows:
\\
\begin{equation}
S=\left(
           \begin{array}{ccccc}
             0 & 0 & 0 & 0 & 1 \\
             1 & 0 & 0 & 0 & 0 \\
             0 & 1 & 0 & 0 & 0 \\
             0 & 0 & 1 & 0 & 0 \\
             0 & 0 & 0 & 1 & 0 \\
           \end{array}
         \right)
\end{equation}
\\
One can see that $A_i$ for $i=1,2\hspace*{1mm}$ are symmetric and
$\sum_{i=0}^{2}A_i =J_5$. Also it can be verified that,
$\{A_i,\hspace*{2mm}i=1,2\}$ is closed under multiplication and
therefore, the set of matrices $A_0, A_1$ and $A_2$ form a symmetric
association scheme.
\\
\\
In view of $A_0, A_1$ and $A_2$ we can write the following cases:
\\
\begin{equation}
A_0,\hspace*{2mm}A_1,\hspace*{2mm}A_2,\hspace*{2mm}A_0
+A_1,\hspace*{2mm}A_0 +A_2,\hspace*{2mm}A_1 +A_2,\hspace*{2mm}A_0
+A_1 +A_2 \hspace*{3cm}
\end{equation}
\\
By examing the number of combinations of 2 cases selected from a set
of the above 7 distinct cases and considering $B_1=S + S^{-1}$ and
$B_2=S^{2} + S^{-2}$ the binary matrix $B=(B_1 \vert B_2)$ is
written as
\\
\begin{equation}
B= \left(
     \begin{array}{ccccccccccc}
       0 & 1 & 0 & 0 & 1 & | & 0 & 0 & 1 & 1 & 0 \\
       1 & 0 & 1 & 0 & 0 & | & 0 & 0 & 0 & 1 & 1 \\
       0 & 1 & 0 & 1 & 0 & | & 1 & 0 & 0 & 0 & 1 \\
       0 & 0 & 1 & 0 & 1 & | & 1 & 1 & 0 & 0 & 0 \\
       1 & 0 & 0 & 1 & 0 & | & 0 & 1 & 1 & 0 & 0 \\
     \end{array}
   \right)
\hspace*{3cm}
\end{equation}
\\
By removing the last row from the binary matrix $B$ we can achieve
$n-k=4$ independent generators. The distance $d$ of the quantum code
is given by the minimum weight of the bitwise OR $\textbf{(a,b)}$ of
all pairs satisfying the symplectic orthogonality condition,
\\
\begin{equation}
B_1 \textbf{b} + B_2 \textbf{a}=0 \hspace*{3cm}
\end{equation}
\\
Let $\textbf{a}=(x_1,x_2,x_3,x_4,x_5)$ and
$\textbf{b}=(y_1,y_2,y_3,y_4,y_5) $. Then by using (3.5), we have
\\
\begin{equation}
\left\{
  \begin{array}{ll}
    x_3 + x_4 +y_2 +y_5 =0 \\
    x_4 + x_5 +y_1 +y_3 =0 \\
    x_1 + x_5 +y_2 +y_4 =0 \\
    x_1 + x_2 +y_3 +y_5 =0
    \end{array}
\right. \hspace*{3cm}
\end{equation}
\\
By using (3.6) we can get the code distance $d$ equal to $3$. Since
the number of independent generators is $n-k=4$, therefore $k=1$,
thus the $[[ 5,1,3 ]]_{2}$ optimal quantum stabilizer code is
constructed. It encodes $k=1$ logical qubit into $n=5$ physical
qubits and protects against an arbitrary single-qubit error. Its
stabilizer consists of $n-k=4$ Pauli operators in table 1.
\\
$$
\begin{tabular}{|c|c|}
  \hline
   Name & Operator\\
  \hline
  $g_1$ &I\hspace*{2mm}X\hspace*{2mm}Z\hspace*{2mm}Z\hspace*{2mm}X \\
  $g_2$ & X\hspace*{2mm}I\hspace*{2mm}X\hspace*{2mm}Z\hspace*{2mm}Z\\
  $g_3$ & Z\hspace*{2mm}X\hspace*{2mm}I\hspace*{2mm}X\hspace*{2mm}Z \\
  $g_4$ & Z\hspace*{2mm}Z\hspace*{2mm}X\hspace*{2mm}I\hspace*{2mm}X \\
  \hline
\end{tabular}
$$
\begin{table}[htb]
\caption{\small{Stabilizer generators for the $[[5,1,3]]_{2}$
code.}} \label{table:2}
\newcommand{\m}{\hphantom{$-$}}
\newcommand{\cc}[1]{\multicolumn{1}{c}{#1}}
\renewcommand{\tabcolsep}{0pc} 
\renewcommand{\arraystretch}{1.} 
\end{table}
\\
\\
Similar to case $m=2$ we obtain quantum stabilizer codes from
$C_\nu,\hspace*{2mm}\nu=6,7, ...\hspace*{1mm}. $ In the case of
$m=3$ from $C_6$ we can write
\\
\begin{equation}
A_0=I_6,\hspace*{4mm}A_1=S^{1}+S^{-1},\hspace*{4mm}A_2=S^{2}+S^{-2},\hspace*{4mm}A_3=S^{3}
 \hspace*{3cm}
\end{equation}
\\
It can be easily seen that $A_i$ for $i=1,2,3$ are symmetric and
$\sum_{i=0}^{3} A_i=J_6$. By choosing $B_1=A_2 +A_3$ and $B_2=A_0
+A_1+A_2$ the binary matrix $B =(B_1 \vert B_2)$ will be in the form
\\
\begin{equation}
B=\left(
    \begin{array}{ccccccccccccc}
      0 & 0 & 1 & 1 & 1 & 0 & | & 1 & 1 & 1 & 0 & 1 & 1 \\
      0 & 0 & 0 & 1 & 1 & 1 & | & 1 & 1 & 1 & 1 & 0 & 1 \\
      1 & 0 & 0 & 0 & 1 & 1 & | & 1 & 1 & 1 & 1 & 1 & 0 \\
      1 & 1 & 0 & 0 & 0 & 1 & | & 0 & 1 & 1 & 1 & 1 & 1 \\
      1 & 1 & 1 & 0 & 0 & 0 & | & 1 & 0 & 1 & 1 & 1 & 1 \\
      0 & 1 & 1 & 1 & 0 & 0 & | & 1 & 1 & 0 & 1 & 1 & 1 \\
    \end{array}
  \right)
\hspace*{3cm}
\end{equation}
\\
By removing the last row from $B$ and constituting the system of
linear equations the analogue of previous case, we can achieve
$d=3$. Since the number of independent generators is $n-k=5$,
therefore the optimal quantum stabilizer code is of length $6$, that
encodes $k=1$ logical qubit, i.e., $[[6,1,3]]_{2}$ is constructed.
This code generated by the $n-k=5$ independent generators in table
$2$.
\\
\\
$$
\begin{tabular}{|c|c|}
  \hline
   Name & Operator\\
  \hline
  $g_1$ &Z\hspace*{2mm}Z\hspace*{2mm}Y\hspace*{2mm}X\hspace*{2mm}Y\hspace*{2mm}Z \\
  $g_2$ & Z\hspace*{2mm}Z\hspace*{2mm}Z\hspace*{2mm}Y\hspace*{2mm}X\hspace*{2mm}Y\\
  $g_3$ & Y\hspace*{2mm}Z\hspace*{2mm}Z\hspace*{2mm}Z\hspace*{2mm}Y\hspace*{2mm}X \\
  $g_4$ & X\hspace*{2mm}Y\hspace*{2mm}Z\hspace*{2mm}Z\hspace*{2mm}Z\hspace*{2mm}Y \\
  $g_5$ & Y\hspace*{2mm}X\hspace*{2mm}Y\hspace*{2mm}Z\hspace*{2mm}Z\hspace*{2mm}Z \\
  \hline
\end{tabular}
$$
\begin{table}[htb]
\caption{\small{Stabilizer generators for the $[[ 6,1,3 ]]_{2}$
code.}} \label{table:2}
\newcommand{\m}{\hphantom{$-$}}
\newcommand{\cc}[1]{\multicolumn{1}{c}{#1}}
\renewcommand{\tabcolsep}{0pc} 
\renewcommand{\arraystretch}{1.} 
\end{table}
\\
\\
To construct a quantum stabilizer code from $C_7$ by using $(2.14)$,
we have
\\
\begin{equation}
A_0=I_7,\hspace*{4mm}A_1=S+S^{-1},\hspace*{4mm}A_2=S^{2}+S^{-2},\hspace*{4mm}A_3=S^{3}+S^{-3}
 \hspace*{2cm}
\end{equation}
\\
One can see that $A_i$ for $i=1,2,3$ are symmetric and
$\sum_{i=0}^{3}A_i=J_7$. Also it can be easily shown that,
$\{A_i,\hspace*{1mm}i=1,2,3\}$ is closed under multiplication and
therefore, the set of matrices $A_0, ... ,A_3$ form a symmetric
association scheme. By choosing $B_1$ and $B_2$ as follows:
\\
\begin{equation}
B_1=A_1,\hspace*{4mm}B_2=A_2+A_3
 \hspace*{3cm}
\end{equation}
\\
We can be seen that $B_1 B_2^{T}+ B_2 B_1^{T}=0$. So all operators
are commute. On the other hand, since
\\
\begin{equation}
B= \left(
  \begin{array}{ccccccccccccccc}
    0 & 1 & 0 & 0 & 0 & 0 & 1 & | & 0 & 0 & 1 & 1 & 1 & 1 & 0 \\
    1 & 0 & 1 & 0 & 0 & 0 & 0 & | & 0 & 0 & 0 & 1 & 1 & 1 & 1 \\
    0 & 1 & 0 & 1 & 0 & 0 & 0 & | & 1 & 0 & 0 & 0 & 1 & 1 & 1 \\
    0 & 0 & 1 & 0 & 1 & 0 & 0 & | & 1 & 1 & 0 & 0 & 0 & 1 & 1 \\
    0 & 0 & 0 & 1 & 0 & 1 & 0 & | & 1 & 1 & 1 & 0 & 0 & 0 & 1 \\
    0 & 0 & 0 & 0 & 1 & 0 & 1 & | & 1 & 1 & 1 & 1 & 0 & 0 & 0 \\
    1 & 0 & 0 & 0 & 0 & 1 & 0 & | & 0 & 1 & 1 & 1 & 1 & 0 & 0 \\
  \end{array}
\right)
 \hspace*{3cm}
\end{equation}
\\
By removing the last row from it by $(3.5)$ the code distance is
$d=3$. And also since the number of independent generators is
$n-k=6$. Therefore, we can obtain the $[[7,1,3]]_{2}$ quantum
stabilizer code. This code generated by $6$ the independent
generators in table $3$.
\\
$$
\begin{tabular}{|c|c|}
  \hline
   Name & Operator\\
  \hline
  $g_1$ & I\hspace*{2mm}X\hspace*{2mm}Z\hspace*{2mm}Z\hspace*{2mm}Z\hspace*{2mm}Z\hspace*{2mm}X\\
  $g_2$ & X\hspace*{2mm}I\hspace*{2mm}X\hspace*{2mm}Z\hspace*{2mm}Z\hspace*{2mm}Z\hspace*{2mm}Z\\
  $g_3$ & Z\hspace*{2mm}X\hspace*{2mm}I\hspace*{2mm}X\hspace*{2mm}Z\hspace*{2mm}Z\hspace*{2mm}Z\\
  $g_4$ & Z\hspace*{2mm}Z\hspace*{2mm}X\hspace*{2mm}I\hspace*{2mm}X\hspace*{2mm}Z\hspace*{2mm}Z\\
  $g_5$ & Z\hspace*{2mm}Z\hspace*{2mm}Z\hspace*{2mm}X\hspace*{2mm}I\hspace*{2mm}X\hspace*{2mm}Z\\
  $g_6$ & Z\hspace*{2mm}Z\hspace*{2mm}Z\hspace*{2mm}Z\hspace*{2mm}X\hspace*{2mm}I\hspace*{2mm}X\\
  \hline
\end{tabular}
$$
\begin{table}[htb]
\caption{\small{Stabilizer generators for the $[[ 7,1,3 ]]_{2}$
code.}} \label{table:2}
\newcommand{\m}{\hphantom{$-$}}
\newcommand{\cc}[1]{\multicolumn{1}{c}{#1}}
\renewcommand{\tabcolsep}{0pc} 
\renewcommand{\arraystretch}{1.} 
\end{table}
\\
Applying (2.12) and (2.14), we can obtain quantum stabilizer codes
from $C_\nu ({\nu}=8,9,...)$.
\\
\\
\textbf{Remark.} A list of binary quantum stabilizer codes from
$C_\nu ({\nu}=8,9,...)$ is given in tables 4 and 5. The first column
shows cyclic groups. The second column shows $B_1$ and $B_2$ in
terms of $A_i$, $i=0,1,...,m$. The third column shows the value of
the length of quantum stabilizer code. The fourth column shows the
value of $n-k$. The fifth column shows a list of the quantum
stabilizer codes. In this table $I_{n}$ is an $n\times n$ unit
matrix and $X$ is an Pauli matrix. Also, we will sometimes use
notation where we omit the tensor signs. For example $A_1I_2I_2$ is
shorthand for $A_1\otimes I_2\otimes I_2$. All the optimal quantum
stabilizer codes, i.e., codes with largest possible $k$ with fixed
$n$ and $d$ constructed in table $4$ lengths labeled by $l$ having
the best parameters known. The highest rate $\frac{k}{n}$ of
$[[n,k,d]]_{2}$ quantum stabilizer codes with minimum distance $d$
is labeled by $u$ in below tables.
\\
\\
$$
\begin{tabular}{|c|p{10.5cm}|c|c|l|}
  \hline
\hline
  Cyclic group  & $B_i(i=1,2)$ & $n$ & $n-k$ & $[[ n,k,d ]]_{2}$\\
  \hline
 $C_{8}$ & $B_1=A_3+A_4$,\hspace*{2mm}$B_2=A_2+A_3$ & $8$ & $6$ & $[[ 8,2,3 ]]_{2}$ \\
 $C_{2}\times C_{4} $ & $B_1=I_{2}A_2+XA_1$,
\hspace*{2mm}$B_2=I_{2}A_1+XA_1+XA_2$ & $8$ & $6$ & $[[ 8,2,3]]_{2}$ \\
 $C_{2}\times C_{2}\times C_{2}$ &
 $B_1=I_2I_2X+XI_2I_2+XI_2X+XXX$,\hspace*{2mm}$B_2=I_2I_2X+I_2XI_2+XXI_2+XXX$ & $^{l}8$ & $5$ & $[[ 8,3,3 ]]_{2}$ \\
 $C_{9}$ & $B_1=A_1+A_2$,\hspace*{2mm}$B_2=A_2+A_4$ & $9$ & $6$ & $[[ 9,3,3 ]]_{2}$ \\
 $C_{3}\times C_{3}$ & $B_1=I_3A_1+SS+S^2S^2$,
\hspace*{2mm}$B_2=I_3A_1+SS^2+S^2S$ & $9$ & $6$ & $[[ 9,3,3 ]]_{2}$ \\
 $C_{10}$ & $B_1=A_2+A_4+A_5$,\hspace*{2mm}$B_2=A_0+A_2+A_3$ & $10$ & $6$ & $[[ 10,4,3 ]]_{2}$ \\
 $C_{10}$ & $B_1=A_4$,\hspace*{2mm}$B_2=A_0+A_3+A_5$ & $10$ & $9$ & $[[ 10,1,4 ]]_{2}$ \\
 $C_{11}$ & $B_1=A_1+A_3+A_4+A_5$,\hspace*{2mm}$B_2=A_2+A_5$ & $11$ & $7$ & $[[ 11,4,3 ]]_{2}$ \\
 $C_{11}$ & $B_1=A_1+A_4+A_5$,\hspace*{2mm}$B_2=A_2+A_5$ & $11$ & $10$ & $[[ 11,1,5 ]]_{2}$ \\
 $C_{12}$ & $B_1=A_2+A_4+A_5+A_6$,\hspace*{2mm}$B_2=A_2+A_3+A_5$ & $^{l}12$ & $6$ & $[[ 12,6,3 ]]_{2}$ \\
 $C_{12}$ & $B_1=A_2+A_4+A_5+A_6$,\hspace*{2mm}$B_2=A_2+A_3+A_5+A_6$ & $12$ & $7$ & $[[ 12,5,3 ]]_{2}$ \\
 $C_{3}\times C_{4} $ & $B_1=I_{12}+I_3A_1+A_1I_4$,
\hspace*{2mm}$B_2=A_1A_1+A_1I_4$ & $12$ & $10$ & $[[ 12,2,3 ]]_{2}$ \\
 $C_{3}\times C_{2}\times C_{2}$ & $B_1=A_1I_2I_2+A_1I_2X+A_1XX$,
\hspace*{2mm}$B_2=I_3XI_2+I_3XX+A_1I_2X$ & $12$ & $8$ & $[[ 12,4,3 ]]_{2}$ \\
 $C_{13}$ & $B_1=A_1+A_3+A_4+A_5$,\hspace*{2mm}$B_2=A_2+A_3+A_5$ & $13$ & $8$ & $[[ 13,5,3 ]]_{2}$ \\
 $C_{13}$ & $B_1=A_1+A_3+A_4+A_5$,\hspace*{2mm}$B_2=A_2+A_3+A_5$ & $13$ & $12$ & $[[ 13,1,5 ]]_{2}$ \\
 $C_{14}$ & $B_1=A_0+A_3+A_4+A_6+A_7$,\hspace*{2mm}$B_2=A_2+A_3+A_5$ & $14$ & $8$ & $[[ 14,6,3 ]]_{2}$ \\
 $C_{14}$ & $B_1=A_0+A_3+A_4+A_6+A_7$,\hspace*{2mm}$B_2=A_2+A_3+A_5$ & $14$ & $11$ & $[[ 14,3,4 ]]_{2}$ \\
 $C_{15}$ & $B_1=A_3+A_4+A_6+A_7$,\hspace*{2mm}$B_2=A_1+A_2+A_3+A_5$ & $15$ & $9$ & $[[ 15,6,3 ]]_{2}$ \\
 $C_{16}$ & $B_1=A_3+A_4+A_6$,\hspace*{2mm}$B_2=A_2+A_3+A_5$ & $16$ & $11$ & $[[ 16,5,3 ]]_{2}$ \\
 $C_{16}$ & $B_1=A_0+A_3+A_4+A_8$,\hspace*{2mm}$B_2=A_0+A_1+A_2+A_5$ & $16$ & $8$ & $[[ 16,8,3 ]]_{2}$ \\
 $C_{2}\times C_{8}$ & $B_1=I_2A_2+XA_2+XA_4+I_2A_3+I_2A_4+XA_1$,
 \hspace*{2mm}$B_2=I_2A_2+XA_3+I_2A_1+I_2A_3+I_2I_8$ & $16$ & $7$ & $[[ 16,9,3 ]]_{2}$ \\
 $C_{2}\times C_{2}\times C_{4}$ & $B_1=I_2I_2A_2+A_1I_2A_1+A_1A_1I_4+I_2A_1I_4+I_2A_1A_1+A_1A_1A_1$,
 \hspace*{0.1mm}$B_2=I_2I_2A_2+A_1I_2A_2+I_2I_2A_1+I_2A_1A_2+I_2A_1I_4+I_2A_1A_1+A_1A_1A_1$ & $16$ & $8$ & $[[ 16,8,3 ]]_{2}$ \\
 $C_{4}\times C_{4}$ & $B_1=I_4A_1+A_1A_1+A_1A_2+A_2A_2$,
 \hspace*{2mm}$B_2=I_4A_2+A_1I_4+A_1A_2+A_2I_4+A_2A_1$ & $16$ & $12$ & $[[ 16,4,3 ]]_{2}$ \\
 $C_{2}\times C_{2}\times C_{2}\times C_{2}$ & $B_1=XI_2I_2X+XI_2XX+XXXX+I_2XXX$,
 \hspace*{2mm}$B_2=I_2I_2I_2X+I_2I_2XI_2+I_2I_2XX+I_2XI_2X+I_2XXI_2+XI_2I_2X+XXI_2I_2+XXXI_2$ & $16$ & $9$ & $[[ 16,7,3 ]]_{2}$ \\
 $C_{17}$ & $B_1=A_3+A_4+A_6+A_7+A_8$,\hspace*{2mm}$B_2=A_2+A_3+A_5$ & $17$ & $10$ & $[[ 17,7,3 ]]_{2}$ \\
 $C_{17}$ & $B_1=A_3+A_4+A_6+A_7+A_8$,\hspace*{2mm}$B_2=A_2+A_3+A_5$ & $17$ & $14$ & $[[ 17,3,4 ]]_{2}$ \\
 $C_{18}$ & $B_1=A_0+A_3+A_4+A_5+A_6$,\hspace*{0.5mm}$B_2=A_3+A_5+A_6+A_7+A_8+A_9$ & $18$ & $10$ & $[[ 18,8,3 ]]_{2}$ \\
 $C_{19}$ & $B_1=A_3+A_4+A_6+A_9$,\hspace*{0.5mm}$B_2=A_3+A_5+A_6+A_7$ & $19$ & $10$ & $[[ 19,9,3 ]]_{2}$ \\
 $C_{20}$ & $B_1=A_3+A_4+A_6+A_9+A_{10}$,\hspace*{0.5mm}$B_2=A_3+A_5+A_6+A_7+A_8$ & $20$ & $8$ & $[[ 20,12,3 ]]_{2}$ \\
 $C_{21}$ & $B_1=A_3+A_4+A_6+A_9+A_{10}$,\hspace*{0.5mm}$B_2=A_3+A_5+A_6+A_7+A_8$ & $21$ & $8$ & $[[ 21,13,3 ]]_{2}$ \\
 $C_{21}$ & $B_1=A_3+A_4+A_6+A_9+A_{10}$,\hspace*{0.5mm}$B_2=A_3+A_5+A_6+A_7+A_8$ & $21$ & $11$ & $[[ 21,10,4 ]]_{2}$ \\
 $C_{21}$ & $B_1=A_3+A_4+A_6+A_9+A_{10}$,\hspace*{0.5mm}$B_2=A_3+A_5+A_6+A_7+A_8$ & $21$ & $12$ & $[[ 21,9,5 ]]_{2}$ \\
 $C_{21}$ & $B_1=A_3+A_4+A_6+A_9+A_{10}$,\hspace*{0.5mm}$B_2=A_3+A_5+A_6+A_7+A_8$ & $^{l}21$ & $16$ & $[[ 21,5,7 ]]_{2}$ \\
\hline \hline
\end{tabular}
$$
\begin{table}[htb]
\caption{\small{Quantum stabilizer codes $[[ n,k,d ]]_{2}$.}}
\label{table:2}
\newcommand{\m}{\hphantom{$-$}}
\newcommand{\cc}[1]{\multicolumn{1}{c}{#1}}
\renewcommand{\tabcolsep}{0pc} 
\renewcommand{\arraystretch}{1.} 
\end{table}
\\
\\
$$
\begin{tabular}{|c|p{10.5cm}|c|c|l|}
\hline
\hline
  Cyclic group  & $B_i(i=1,2)$ & $n$ & $n-k$ & $[[ n,k,d ]]_{2}$\\
  \hline
 $C_{25}$ & $B_1=A_3+A_4+A_6+A_9+A_{10}$,\hspace*{0.5mm}$B_2=A_3+A_5+A_6+A_7+A_8$ & $25$ & $8$ & $[[ 25,17,3 ]]_{2}$ \\
 $C_{25}$ & $B_1=A_3+A_4+A_6+A_9+A_{10}$,\hspace*{0.5mm}$B_2=A_3+A_5+A_6+A_7+A_8$ & $25$ & $12$ & $[[ 25,13,4 ]]_{2}$ \\
 $C_{30}$ & $B_1=A_3+A_4+A_6+A_9+A_{10}+A_{14}$,\hspace*{0.5mm}$B_2=A_3+A_5+A_6+A_7+A_8+A_{13}+A_{15}$
  & $30$ & $8$ & $[[ 30,22,3 ]]_{2}$ \\
 $C_{30}$ & $B_1=A_3+A_4+A_6+A_9+A_{10}+A_{14}$,\hspace*{0.5mm}$B_2=A_3+A_5+A_6+A_7+A_8+A_{13}+A_{15}$
  & $30$ & $18$ & $[[ 30,12,5 ]]_{2}$ \\
 $C_{40}$ & $B_1=A_3+A_4+A_6+A_9+A_{10}+A_{12}+A_{14}+A_{15}+A_{16}+A_{18}+A_{19}$,
 \hspace*{0.5mm}$B_2=A_3+A_5+A_6+A_7+A_8+A_{12}+A_{15}+A_{16}+A_{17}+A_{18}+A_{20}$
  & $40$ & $10$ & $^{u}[[ 40,30,3 ]]_{2}$ \\
 $C_{40}$ & $B_1=A_3+A_4+A_6+A_9+A_{10}+A_{12}+A_{14}+A_{15}+A_{16}+A_{18}+A_{19}$,
 \hspace*{0.5mm}$B_2=A_3+A_5+A_6+A_7+A_8+A_{12}+A_{15}+A_{16}+A_{17}+A_{18}+A_{20}$
  & $40$ & $14$ & $^{u}[[ 40,26,5 ]]_{2}$ \\
   $C_{40}$ & $B_1=A_3+A_4+A_6+A_9+A_{10}+A_{12}+A_{14}+A_{15}+A_{16}+A_{18}+A_{19}$,
 \hspace*{0.5mm}$B_2=A_3+A_5+A_6+A_7+A_8+A_{12}+A_{15}+A_{16}+A_{17}+A_{18}+A_{20}$
  & $^{l}40$ & $19$ & $^{u}[[ 40,21,7 ]]_{2}$ \\
  \hline
\hline
\end{tabular}
$$
\begin{table}[htb]
\caption{\small{Quantum stabilizer codes $[[ n,k,d ]]_{2}$.}}
\label{table:2}
\newcommand{\m}{\hphantom{$-$}}
\newcommand{\cc}[1]{\multicolumn{1}{c}{#1}}
\renewcommand{\tabcolsep}{0pc} 
\renewcommand{\arraystretch}{1.} 
\end{table}
\subsection{\hspace*{-.3cm}construction of quantum stabilizer codes
of distances five and seven from Abelian group association schemes}
We can extend the stabilizers of the codes from section $3$ to get
distances five and seven codes. The parameters of these codes with
$d=5,7$ will be $[[ n,k,d ]]_{2}$. In the case of $m=5$ from
$C_{11}$ we can write
\\
\begin{equation}
A_0=I_{11},\hspace*{0.5mm}A_1=S^{1}+S^{-1},\hspace*{0.5mm}A_2=S^{2}+S^{-2},\hspace*{0.5mm}A_3=S^{3}+S^{-3},\hspace*{0.5mm}
A_4=S^{4}+S^{-4},\hspace*{0.5mm}A_5=S^{5}+S^{-5}
\end{equation}
\\
where $S$ is an $11\times 11$ circulant matrix with period $11$
$(S^{11}=I_{11})$. One can easily see that the above adjacency
matrices for $i=1,...,5$ are symmetric and
$\sum_{i=0}^{5}A_i=J_{11}$. Also, the set of matrices $A_0, ...
,A_5$ form a symmetric association scheme. By choosing
$B_1=A_1+A_4+A_5$ and $B_2=A_2 +A_5$ the binary matrix $B =(B_1
\vert B_2)$ will be in the form
\\
\begin{equation}
B=\left(
    \begin{array}{ccccccccccccccccccccccc}
      0 & 1 & 0 & 0 & 1 & 1 & 1 & 1 & 0 & 0 & 1 & | & 0 & 0 & 1 & 0 & 0 & 1 & 1 & 0 & 0 & 1 & 0\\
      1 & 0 & 1 & 0 & 0 & 1 & 1 & 1 & 1 & 0 & 0 & | & 0 & 0 & 0 & 1 & 0 & 0 & 1 & 1 & 0 & 0 & 1\\
      0 & 1 & 0 & 1 & 0 & 0 & 1 & 1 & 1 & 1 & 0 & | & 1 & 0 & 0 & 0 & 1 & 0 & 0 & 1 & 1 & 0 & 0\\
      0 & 0 & 1 & 0 & 1 & 0 & 0 & 1 & 1 & 1 & 1 & | & 0 & 1 & 0 & 0 & 0 & 1 & 0 & 0 & 1 & 1 & 0\\
      1 & 0 & 0 & 1 & 0 & 1 & 0 & 0 & 1 & 1 & 1 & | & 0 & 0 & 1 & 0 & 0 & 0 & 1 & 0 & 0 & 1 & 1\\
      1 & 1 & 0 & 0 & 1 & 0 & 1 & 0 & 0 & 1 & 1 & | & 1 & 0 & 0 & 1 & 0 & 0 & 0 & 1 & 0 & 0 & 1\\
      1 & 1 & 1 & 0 & 0 & 1 & 0 & 1 & 0 & 0 & 1 & | & 1 & 1 & 0 & 0 & 1 & 0 & 0 & 0 & 1 & 0 & 0\\
      1 & 1 & 1 & 1 & 0 & 0 & 1 & 0 & 1 & 0 & 0 & | & 0 & 1 & 1 & 0 & 0 & 1 & 0 & 0 & 0 & 1 & 0\\
      0 & 1 & 1 & 1 & 1 & 0 & 0 & 1 & 0 & 1 & 0 & | & 0 & 0 & 1 & 1 & 0 & 0 & 1 & 0 & 0 & 0 & 1\\
      0 & 0 & 1 & 1 & 1 & 1 & 0 & 0 & 1 & 0 & 1 & | & 1 & 0 & 0 & 1 & 1 & 0 & 0 & 1 & 0 & 0 & 0\\
      1 & 0 & 0 & 1 & 1 & 1 & 1 & 0 & 0 & 1 & 0 & | & 0 & 1 & 0 & 0 & 1 & 1 & 0 & 0 & 1 & 0 & 0\\
    \end{array}
  \right)
\end{equation}
\\
By removing the last row from $B$ and by considering
$\textbf{a}=(x_{01},...,x_{11})$ and
$\textbf{b}=(y_{01},...,y_{11})$, in view of (3.5) we can achieve
$d=5$.
\\
Since the number of independent generators is $n-k=10$, therefore
the quantum stabilizer code is of length $11$, that encodes $k=1$
logical qubit, i.e., $[[11,1,5]]_{2}$ is constructed. This code
generated by the $n-k=10$ independent generators in table $6$.
\\
\\
$$
\begin{tabular}{|c|c|}
  \hline
   Name & Operator\\
  \hline
  $g_1$ &I\hspace*{2mm}X\hspace*{2mm}Z\hspace*{2mm}I\hspace*{2mm}X\hspace*{2mm}Y\hspace*{2mm}Y\hspace*{2mm}X\hspace*{2mm}I\hspace*{2mm}Z\hspace*{2mm}X\\
  $g_2$ &X\hspace*{2mm}I\hspace*{2mm}X\hspace*{2mm}Z\hspace*{2mm}I\hspace*{2mm}X\hspace*{2mm}Y\hspace*{2mm}Y\hspace*{2mm}X\hspace*{2mm}I\hspace*{2mm}Z\\
  $g_3$ &Z\hspace*{2mm}X\hspace*{2mm}I\hspace*{2mm}X\hspace*{2mm}Z\hspace*{2mm}I\hspace*{2mm}X\hspace*{2mm}Y\hspace*{2mm}Y\hspace*{2mm}X\hspace*{2mm}I\\
  $g_4$ &I\hspace*{2mm}Z\hspace*{2mm}X\hspace*{2mm}I\hspace*{2mm}X\hspace*{2mm}Z\hspace*{2mm}I\hspace*{2mm}X\hspace*{2mm}Y\hspace*{2mm}Y\hspace*{2mm}X\\
  $g_5$ &X\hspace*{2mm}I\hspace*{2mm}Z\hspace*{2mm}X\hspace*{2mm}I\hspace*{2mm}X\hspace*{2mm}Z\hspace*{2mm}I\hspace*{2mm}X\hspace*{2mm}Y\hspace*{2mm}Y\\
  $g_6$ &Y\hspace*{2mm}X\hspace*{2mm}I\hspace*{2mm}Z\hspace*{2mm}X\hspace*{2mm}I\hspace*{2mm}X\hspace*{2mm}Z\hspace*{2mm}I\hspace*{2mm}X\hspace*{2mm}Y\\
  $g_8$ &X\hspace*{2mm}Y\hspace*{2mm}Y\hspace*{2mm}X\hspace*{2mm}I\hspace*{2mm}Z\hspace*{2mm}X\hspace*{2mm}I\hspace*{2mm}X\hspace*{2mm}Z\hspace*{2mm}I\\
  $g_9$ &I\hspace*{2mm}X\hspace*{2mm}Y\hspace*{2mm}Y\hspace*{2mm}X\hspace*{2mm}I\hspace*{2mm}Z\hspace*{2mm}X\hspace*{2mm}I\hspace*{2mm}X\hspace*{2mm}Z\\
  $g_{10}$ &Z\hspace*{2mm}I\hspace*{2mm}X\hspace*{2mm}Y\hspace*{2mm}Y\hspace*{2mm}X\hspace*{2mm}I\hspace*{2mm}Z\hspace*{2mm}X\hspace*{2mm}I\hspace*{2mm}X\\
  \hline
\end{tabular}
$$
\begin{table}[htb]
\caption{\small{Stabilizer generators for the $[[ 11,1,5 ]]_{2}$
code.}} \label{table:2}
\newcommand{\m}{\hphantom{$-$}}
\newcommand{\cc}[1]{\multicolumn{1}{c}{#1}}
\renewcommand{\tabcolsep}{0pc} 
\renewcommand{\arraystretch}{1.} 
\end{table}
\\
\\
For construction of distance five quantum stabilizer code from
$C_{13}$ by using $(2.14)$, we have
\\
\\
\begin{equation}
A_0=I_{13},A_1=S+S^{-1},A_2=S^{2}+S^{-2},A_3=S^{3}+S^{-3},
 A_4=S^{4}+S^{-4},A_5=S^{5}+S^{-5},A_6=S^{6}+S^{-6}
\end{equation}
\\
One can see that $A_i$ for $i=1,...,6$ are symmetric and
$\sum_{i=0}^{6}A_i=J_{13}$. Also it can be easily shown that,
$\{A_i,\hspace*{1mm}i=1,...,6\}$ is closed under multiplication and
therefore, the set of matrices $A_0, ... ,A_6$ form a symmetric
association scheme. By choosing $B_1$ and $B_2$ as follows:
\\
\begin{equation}
B_1=A_1+A_3+A_4+A_5,\hspace*{4mm}B_2=A_2+A_3+A_5
 \hspace*{3cm}
\end{equation}
\\
We can be seen that $B_1 B_2^{T}+ B_2 B_1^{T}=0$. So all operators
are commute. On the other hand, since
\\
\begin{equation}
B=\left(
    \begin{array}{ccccccccccccccccccccccccccc}
      0 & 1 & 0 & 1 & 1 & 1 & 0 & 0 & 1 & 1 & 1 & 0 & 1 \hspace*{2mm} |\hspace*{2mm} 0 & 0 & 1 & 1 & 0 & 1 & 0 & 0 & 1 & 0 & 1 & 1 & 0\\
      1 & 0 & 1 & 0 & 1 & 1 & 1 & 0 & 0 & 1 & 1 & 1 & 0 \hspace*{2mm} |\hspace*{2mm} 0 & 0 & 0 & 1 & 1 & 0 & 1 & 0 & 0 & 1 & 0 & 1 & 1\\
      0 & 1 & 0 & 1 & 0 & 1 & 1 & 1 & 0 & 0 & 1 & 1 & 1 \hspace*{2mm} |\hspace*{2mm} 1 & 0 & 0 & 0 & 1 & 1 & 0 & 1 & 0 & 0 & 1 & 0 & 1\\
      1 & 0 & 1 & 0 & 1 & 0 & 1 & 1 & 1 & 0 & 0 & 1 & 1 \hspace*{2mm} |\hspace*{2mm} 1 & 1 & 0 & 0 & 0 & 1 & 1 & 0 & 1 & 0 & 0 & 1 & 0\\
      1 & 1 & 0 & 1 & 0 & 1 & 0 & 1 & 1 & 1 & 0 & 0 & 1 \hspace*{2mm} |\hspace*{2mm} 0 & 1 & 1 & 0 & 0 & 0 & 1 & 1 & 0 & 1 & 0 & 0 & 1\\
      1 & 1 & 1 & 0 & 1 & 0 & 1 & 0 & 1 & 1 & 1 & 0 & 0 \hspace*{2mm} |\hspace*{2mm} 1 & 0 & 1 & 1 & 0 & 0 & 0 & 1 & 1 & 0 & 1 & 0 & 0\\
      0 & 1 & 1 & 1 & 0 & 1 & 0 & 1 & 0 & 1 & 1 & 1 & 0 \hspace*{2mm} |\hspace*{2mm} 0 & 1 & 0 & 1 & 1 & 0 & 0 & 0 & 1 & 1 & 0 & 1 & 0\\
      0 & 0 & 1 & 1 & 1 & 0 & 1 & 0 & 1 & 0 & 1 & 1 & 1 \hspace*{2mm} |\hspace*{2mm} 0 & 0 & 1 & 0 & 1 & 1 & 0 & 0 & 0 & 1 & 1 & 0 & 1\\
      1 & 0 & 0 & 1 & 1 & 1 & 0 & 1 & 0 & 1 & 0 & 1 & 1 \hspace*{2mm} |\hspace*{2mm} 1 & 0 & 0 & 1 & 0 & 1 & 1 & 0 & 0 & 0 & 1 & 1 & 0\\
      1 & 1 & 0 & 0 & 1 & 1 & 1 & 0 & 1 & 0 & 1 & 0 & 1 \hspace*{2mm} |\hspace*{2mm} 0 & 1 & 0 & 0 & 1 & 0 & 1 & 1 & 0 & 0 & 0 & 1 & 1\\
      1 & 1 & 1 & 0 & 0 & 1 & 1 & 1 & 0 & 1 & 0 & 1 & 0 \hspace*{2mm} |\hspace*{2mm} 1 & 0 & 1 & 0 & 0 & 1 & 0 & 1 & 1 & 0 & 0 & 0 & 1\\
      0 & 1 & 1 & 1 & 0 & 0 & 1 & 1 & 1 & 0 & 1 & 0 & 1 \hspace*{2mm} |\hspace*{2mm} 1 & 1 & 0 & 1 & 0 & 0 & 1 & 0 & 1 & 1 & 0 & 0 & 0\\
      1 & 0 & 1 & 1 & 1 & 0 & 0 & 1 & 1 & 1 & 0 & 1 & 0 \hspace*{2mm} |\hspace*{2mm} 0 & 1 & 1 & 0 & 1 & 0 & 0 & 1 & 0 & 1 & 1 & 0 & 0\\
 \end{array}
  \right)
\end{equation}
\\
By removing the last row from $B$ and by considering
$\textbf{a}=(x_{01},...,x_{13})$ and
$\textbf{b}=(y_{01},...,y_{13})$, in view of (3.5) we can achieve
$d=5$.
\\
Since the number of independent generators is $n-k=12$, therefore
the quantum stabilizer code is of length $13$, that encodes $k=1$
logical qubit, i.e., $[[13,1,5]]_{2}$ is constructed. This code
generated by the $n-k=12$ independent generators in table $7$.
\\
$$
\begin{tabular}{|c|c|}
  \hline
   Name & Operator\\
  \hline
  $g_1$ &I\hspace*{2mm}X\hspace*{2mm}Z\hspace*{2mm}Y\hspace*{2mm}X\hspace*{2mm}Y\hspace*{2mm}I\hspace*{2mm}I\hspace*{2mm}Y\hspace*{2mm}X\hspace*{2mm}Y\hspace*{2mm}Z\hspace*{2mm}X\\
  $g_2$ &X\hspace*{2mm}I\hspace*{2mm}X\hspace*{2mm}Z\hspace*{2mm}Y\hspace*{2mm}X\hspace*{2mm}Y\hspace*{2mm}I\hspace*{2mm}I\hspace*{2mm}Y\hspace*{2mm}X\hspace*{2mm}Y\hspace*{2mm}Z\\
  $g_3$ &Z\hspace*{2mm}X\hspace*{2mm}I\hspace*{2mm}X\hspace*{2mm}Z\hspace*{2mm}Y\hspace*{2mm}X\hspace*{2mm}Y\hspace*{2mm}I\hspace*{2mm}I\hspace*{2mm}Y\hspace*{2mm}X\hspace*{2mm}Y\\
  $g_4$ &Y\hspace*{2mm}Z\hspace*{2mm}X\hspace*{2mm}I\hspace*{2mm}X\hspace*{2mm}Z\hspace*{2mm}Y\hspace*{2mm}X\hspace*{2mm}Y\hspace*{2mm}I\hspace*{2mm}I\hspace*{2mm}Y\hspace*{2mm}X\\
  $g_5$ &X\hspace*{2mm}Y\hspace*{2mm}Z\hspace*{2mm}X\hspace*{2mm}I\hspace*{2mm}X\hspace*{2mm}Z\hspace*{2mm}Y\hspace*{2mm}X\hspace*{2mm}Y\hspace*{2mm}I\hspace*{2mm}I\hspace*{2mm}Y\\
  $g_6$ &Y\hspace*{2mm}X\hspace*{2mm}Y\hspace*{2mm}Z\hspace*{2mm}X\hspace*{2mm}I\hspace*{2mm}X\hspace*{2mm}Z\hspace*{2mm}Y\hspace*{2mm}X\hspace*{2mm}Y\hspace*{2mm}I\hspace*{2mm}I\\
  $g_7$ &I\hspace*{2mm}Y\hspace*{2mm}X\hspace*{2mm}Y\hspace*{2mm}Z\hspace*{2mm}X\hspace*{2mm}I\hspace*{2mm}X\hspace*{2mm}Z\hspace*{2mm}Y\hspace*{2mm}X\hspace*{2mm}Y\hspace*{2mm}I\\
  $g_8$ &I\hspace*{2mm}I\hspace*{2mm}Y\hspace*{2mm}X\hspace*{2mm}Y\hspace*{2mm}Z\hspace*{2mm}X\hspace*{2mm}I\hspace*{2mm}X\hspace*{2mm}Z\hspace*{2mm}Y\hspace*{2mm}X\hspace*{2mm}Y\\
  $g_9$ &Y\hspace*{2mm}I\hspace*{2mm}I\hspace*{2mm}Y\hspace*{2mm}X\hspace*{2mm}Y\hspace*{2mm}Z\hspace*{2mm}X\hspace*{2mm}I\hspace*{2mm}X\hspace*{2mm}Z\hspace*{2mm}Y\hspace*{2mm}X\\
  $g_{10}$ &X\hspace*{2mm}Y\hspace*{2mm}I\hspace*{2mm}I\hspace*{2mm}Y\hspace*{2mm}X\hspace*{2mm}Y\hspace*{2mm}Z\hspace*{2mm}X\hspace*{2mm}I\hspace*{2mm}X\hspace*{2mm}Z\hspace*{2mm}Y\\
  $g_{11}$ &Y\hspace*{2mm}X\hspace*{2mm}Y\hspace*{2mm}I\hspace*{2mm}I\hspace*{2mm}Y\hspace*{2mm}X\hspace*{2mm}Y\hspace*{2mm}Z\hspace*{2mm}X\hspace*{2mm}I\hspace*{2mm}X\hspace*{2mm}Z\\
  $g_{12}$ &Z\hspace*{2mm}Y\hspace*{2mm}X\hspace*{2mm}Y\hspace*{2mm}I\hspace*{2mm}I\hspace*{2mm}Y\hspace*{2mm}X\hspace*{2mm}Y\hspace*{2mm}Z\hspace*{2mm}X\hspace*{2mm}I\hspace*{2mm}X\\
  \hline
\end{tabular}
$$
\begin{table}[htb]
\caption{\small{Stabilizer generators for the $[[ 13,1,5 ]]_{2}$
code.}} \label{table:2}
\newcommand{\m}{\hphantom{$-$}}
\newcommand{\cc}[1]{\multicolumn{1}{c}{#1}}
\renewcommand{\tabcolsep}{0pc} 
\renewcommand{\arraystretch}{1.} 
\end{table}
\\
For the construction of distance five quantum stabilizer code from
$C_{21}$ we choose $B_1$ and $B_2$ as follows:
\\
\begin{equation}
B_1=A_3+A_4+A_6+A_9+A_{10},\hspace*{4mm}B_2=A_3+A_5+A_6+A_7+A_8
\end{equation}
\\
We can be seen that $B_1 B_2^{T}+ B_2 B_1^{T}=0$. So all operators
are commute. By removing the last nine rows from $B=(B_1 \vert B_2)$
and by considering $\textbf{a}=(x_{01},...,x_{21})$ and
$\textbf{b}=(y_{01},...,y_{21})$, in view of (3.5) we can achieve
$d=7$.
\\
Since the number of independent generators is $n-k=16$, therefore
the optimal quantum stabilizer code is of length $21$, that encodes
$k=5$ logical qubit, i.e., $[[21,5,7]]_{2}$ is constructed. This
code generated by the $n-k=16$ independent generators in table $8$.
The rate $\frac{k}{n}$ of $[[21,5,7]]_{2}$ code is $0.238$.
\\
$$
\begin{tabular}{|c|c|}
  \hline
   Name & Operator\\
  \hline
  $g_1$ &I\hspace*{2mm}I\hspace*{2mm}I\hspace*{2mm}Y\hspace*{2mm}X\hspace*{2mm}Z\hspace*{2mm}Y\hspace*{2mm}Z\hspace*{2mm}Z\hspace*{2mm}X\hspace*{2mm}X\hspace*{2mm}X\hspace*{2mm}X\hspace*{2mm}Z\hspace*{2mm}Z\hspace*{2mm}Y\hspace*{2mm}Z\hspace*{2mm}X\hspace*{2mm}Y\hspace*{2mm}I\hspace*{2mm}I\\
  $g_2$ &I\hspace*{2mm}I\hspace*{2mm}I\hspace*{2mm}I\hspace*{2mm}Y\hspace*{2mm}X\hspace*{2mm}Z\hspace*{2mm}Y\hspace*{2mm}Z\hspace*{2mm}Z\hspace*{2mm}X\hspace*{2mm}X\hspace*{2mm}X\hspace*{2mm}X\hspace*{2mm}Z\hspace*{2mm}Z\hspace*{2mm}Y\hspace*{2mm}Z\hspace*{2mm}X\hspace*{2mm}Y\hspace*{2mm}I\\
  $g_3$ &I\hspace*{2mm}I\hspace*{2mm}I\hspace*{2mm}I\hspace*{2mm}I\hspace*{2mm}Y\hspace*{2mm}X\hspace*{2mm}Z\hspace*{2mm}Y\hspace*{2mm}Z\hspace*{2mm}Z\hspace*{2mm}X\hspace*{2mm}X\hspace*{2mm}X\hspace*{2mm}X\hspace*{2mm}Z\hspace*{2mm}Z\hspace*{2mm}Y\hspace*{2mm}Z\hspace*{2mm}X\hspace*{2mm}Y\\
  $g_4$ &Y\hspace*{2mm}I\hspace*{2mm}I\hspace*{2mm}I\hspace*{2mm}I\hspace*{2mm}I\hspace*{2mm}Y\hspace*{2mm}X\hspace*{2mm}Z\hspace*{2mm}Y\hspace*{2mm}Z\hspace*{2mm}Z\hspace*{2mm}X\hspace*{2mm}X\hspace*{2mm}X\hspace*{2mm}X\hspace*{2mm}Z\hspace*{2mm}Z\hspace*{2mm}Y\hspace*{2mm}Z\hspace*{2mm}X\\
  $g_5$ &X\hspace*{2mm}Y\hspace*{2mm}I\hspace*{2mm}I\hspace*{2mm}I\hspace*{2mm}I\hspace*{2mm}I\hspace*{2mm}Y\hspace*{2mm}X\hspace*{2mm}Z\hspace*{2mm}Y\hspace*{2mm}Z\hspace*{2mm}Z\hspace*{2mm}X\hspace*{2mm}X\hspace*{2mm}X\hspace*{2mm}X\hspace*{2mm}Z\hspace*{2mm}Z\hspace*{2mm}Y\hspace*{2mm}Z\\
  $g_6$ &Z\hspace*{2mm}X\hspace*{2mm}Y\hspace*{2mm}I\hspace*{2mm}I\hspace*{2mm}I\hspace*{2mm}I\hspace*{2mm}I\hspace*{2mm}Y\hspace*{2mm}X\hspace*{2mm}Z\hspace*{2mm}Y\hspace*{2mm}Z\hspace*{2mm}Z\hspace*{2mm}X\hspace*{2mm}X\hspace*{2mm}X\hspace*{2mm}X\hspace*{2mm}Z\hspace*{2mm}Z\hspace*{2mm}Y\\
  $g_7$ &Y\hspace*{2mm}Z\hspace*{2mm}X\hspace*{2mm}Y\hspace*{2mm}I\hspace*{2mm}I\hspace*{2mm}I\hspace*{2mm}I\hspace*{2mm}I\hspace*{2mm}Y\hspace*{2mm}X\hspace*{2mm}Z\hspace*{2mm}Y\hspace*{2mm}Z\hspace*{2mm}Z\hspace*{2mm}X\hspace*{2mm}X\hspace*{2mm}X\hspace*{2mm}X\hspace*{2mm}Z\hspace*{2mm}Z\\
  $g_8$ &Z\hspace*{2mm}Y\hspace*{2mm}Z\hspace*{2mm}X\hspace*{2mm}Y\hspace*{2mm}I\hspace*{2mm}I\hspace*{2mm}I\hspace*{2mm}I\hspace*{2mm}I\hspace*{2mm}Y\hspace*{2mm}X\hspace*{2mm}Z\hspace*{2mm}Y\hspace*{2mm}Z\hspace*{2mm}Z\hspace*{2mm}X\hspace*{2mm}X\hspace*{2mm}X\hspace*{2mm}X\hspace*{2mm}Z\\
  $g_9$ &Z\hspace*{2mm}Z\hspace*{2mm}Y\hspace*{2mm}Z\hspace*{2mm}X\hspace*{2mm}Y\hspace*{2mm}I\hspace*{2mm}I\hspace*{2mm}I\hspace*{2mm}I\hspace*{2mm}I\hspace*{2mm}Y\hspace*{2mm}X\hspace*{2mm}Z\hspace*{2mm}Y\hspace*{2mm}Z\hspace*{2mm}Z\hspace*{2mm}X\hspace*{2mm}X\hspace*{2mm}X\hspace*{2mm}X\\
  $g_{10}$ &X\hspace*{2mm}Z\hspace*{2mm}Z\hspace*{2mm}Y\hspace*{2mm}Z\hspace*{2mm}X\hspace*{2mm}Y\hspace*{2mm}I\hspace*{2mm}I\hspace*{2mm}I\hspace*{2mm}I\hspace*{2mm}I\hspace*{2mm}Y\hspace*{2mm}X\hspace*{2mm}Z\hspace*{2mm}Y\hspace*{2mm}Z\hspace*{2mm}Z\hspace*{2mm}X\hspace*{2mm}X\hspace*{2mm}X\\
  $g_{11}$ &X\hspace*{2mm}X\hspace*{2mm}Z\hspace*{2mm}Z\hspace*{2mm}Y\hspace*{2mm}Z\hspace*{2mm}X\hspace*{2mm}Y\hspace*{2mm}I\hspace*{2mm}I\hspace*{2mm}I\hspace*{2mm}I\hspace*{2mm}I\hspace*{2mm}Y\hspace*{2mm}X\hspace*{2mm}Z\hspace*{2mm}Y\hspace*{2mm}Z\hspace*{2mm}Z\hspace*{2mm}X\hspace*{2mm}X\\
  $g_{12}$ &X\hspace*{2mm}X\hspace*{2mm}X\hspace*{2mm}Z\hspace*{2mm}Z\hspace*{2mm}Y\hspace*{2mm}Z\hspace*{2mm}X\hspace*{2mm}Y\hspace*{2mm}I\hspace*{2mm}I\hspace*{2mm}I\hspace*{2mm}I\hspace*{2mm}I\hspace*{2mm}Y\hspace*{2mm}X\hspace*{2mm}Z\hspace*{2mm}Y\hspace*{2mm}Z\hspace*{2mm}Z\hspace*{2mm}X\\
  $g_{13}$ &X\hspace*{2mm}X\hspace*{2mm}X\hspace*{2mm}X\hspace*{2mm}Z\hspace*{2mm}Z\hspace*{2mm}Y\hspace*{2mm}Z\hspace*{2mm}X\hspace*{2mm}Y\hspace*{2mm}I\hspace*{2mm}I\hspace*{2mm}I\hspace*{2mm}I\hspace*{2mm}I\hspace*{2mm}Y\hspace*{2mm}X\hspace*{2mm}Z\hspace*{2mm}Y\hspace*{2mm}Z\hspace*{2mm}Z\\
  $g_{14}$ &Z\hspace*{2mm}X\hspace*{2mm}X\hspace*{2mm}X\hspace*{2mm}X\hspace*{2mm}Z\hspace*{2mm}Z\hspace*{2mm}Y\hspace*{2mm}Z\hspace*{2mm}X\hspace*{2mm}Y\hspace*{2mm}I\hspace*{2mm}I\hspace*{2mm}I\hspace*{2mm}I\hspace*{2mm}I\hspace*{2mm}Y\hspace*{2mm}X\hspace*{2mm}Z\hspace*{2mm}Y\hspace*{2mm}Z\\
  $g_{15}$ &Z\hspace*{2mm}Z\hspace*{2mm}X\hspace*{2mm}X\hspace*{2mm}X\hspace*{2mm}X\hspace*{2mm}Z\hspace*{2mm}Z\hspace*{2mm}Y\hspace*{2mm}Z\hspace*{2mm}X\hspace*{2mm}Y\hspace*{2mm}I\hspace*{2mm}I\hspace*{2mm}I\hspace*{2mm}I\hspace*{2mm}I\hspace*{2mm}Y\hspace*{2mm}X\hspace*{2mm}Z\hspace*{2mm}Y\\
  $g_{16}$ &Y\hspace*{2mm}Z\hspace*{2mm}Z\hspace*{2mm}X\hspace*{2mm}X\hspace*{2mm}X\hspace*{2mm}X\hspace*{2mm}Z\hspace*{2mm}Z\hspace*{2mm}Y\hspace*{2mm}Z\hspace*{2mm}X\hspace*{2mm}Y\hspace*{2mm}I\hspace*{2mm}I\hspace*{2mm}I\hspace*{2mm}I\hspace*{2mm}I\hspace*{2mm}Y\hspace*{2mm}X\hspace*{2mm}Z\\
  \hline
\end{tabular}
$$
\begin{table}[htb]
\caption{\small{Stabilizer generators for the $[[ 21,5,7 ]]_{2}$
code.}} \label{table:2}
\newcommand{\m}{\hphantom{$-$}}
\newcommand{\cc}[1]{\multicolumn{1}{c}{#1}}
\renewcommand{\tabcolsep}{0pc} 
\renewcommand{\arraystretch}{1.} 
\end{table}
\\
\\
\section{\hspace*{-.3cm}Construction of stabilizer codes from non-Abelian group association schemes}
The construction of binary quantum stabilizer codes based on the
non-Abelian group association schemes as in the case of Abelian
group association schemes. To do so, we choose a binary matrix
$A=(A_1 \vert A_2)$, such that by removing arbitrarily row or rows
from $A$ we can obtain $n-k$ independent generators. After finding
the code distance by $n-k$ independent generators we can then
determine the parameters of the associated code.
\\
\\
Consider the group $U_{6n}$, as is presented in section 2.5. By
setting $n=2$ in view of (2.27), we have
\\
\begin{align}
A_0 &=I_{12},\nonumber\\
A_1 &=[a]^2,\nonumber\\
A_2 &=[b]+[b]^2, \\
A_3 &=[b][a]^2+[b]^2[a]^2,\nonumber\\
A_4 &=[a]+[b][a]+[b]^2[a],\nonumber\\
A_{5} &=[a]^3+[b][a]^3+[b]^2[a]^3 \nonumber
 \hspace*{3cm}
\end{align}
\\
\\
One can see that $\sum_{i=0}^{5}A_i =J_{12}$, $A_i^{T}\in
\{A_0,A_1,...,A_5\}$ for $0\leq i \leq 5$, and $A_iA_j$ is a linear
combination of $A_0,A_1,...,A_5$ for $0\leq i,j\leq 5$  . Also it
can be verified that, $\{A_i,\hspace*{2mm}i=1,...,5\}$ is closed
under multiplication and therefore, the set of matrices
$A_0,A_1,...,A_5$ form an association scheme with $5$ classes.
\\
\\
By examing the number of combinations of 2 cases selected from a set
of 63 distinct cases and considering $B_1=A_2$ and $B_2=A_3+A_5$ the
binary matrix $B=(B_1 \vert B_2)$ is written as
\\
\begin{equation}
B=\left(
    \begin{array}{cccccccccccccccccccccccc}
      0 & 0 & 0 & 0 & 1 & 0 & 0 & 0 & 1 & 0 & 0 & 0 \hspace*{2mm} |\hspace*{2mm} 0 & 1 & 0 & 0 & 0 & 1 & 1 & 0 & 0 & 1 & 1 & 0 \\
      0 & 0 & 0 & 0 & 0 & 1 & 0 & 0 & 0 & 1 & 0 & 0 \hspace*{2mm} |\hspace*{2mm} 0 & 0 & 1 & 0 & 0 & 0 & 1 & 1 & 0 & 0 & 1 & 1 \\
      0 & 0 & 0 & 0 & 0 & 0 & 1 & 0 & 0 & 0 & 1 & 0 \hspace*{2mm} |\hspace*{2mm} 0 & 0 & 0 & 1 & 1 & 0 & 0 & 1 & 1 & 0 & 0 & 1 \\
      0 & 0 & 0 & 0 & 0 & 0 & 0 & 1 & 0 & 0 & 0 & 1 \hspace*{2mm} |\hspace*{2mm} 1 & 0 & 0 & 0 & 1 & 1 & 0 & 0 & 1 & 1 & 0 & 0 \\
      1 & 0 & 0 & 0 & 0 & 0 & 0 & 0 & 1 & 0 & 0 & 0 \hspace*{2mm} |\hspace*{2mm} 0 & 1 & 1 & 0 & 0 & 1 & 0 & 0 & 0 & 1 & 1 & 0 \\
      0 & 1 & 0 & 0 & 0 & 0 & 0 & 0 & 0 & 1 & 0 & 0 \hspace*{2mm} |\hspace*{2mm} 0 & 0 & 1 & 1 & 0 & 0 & 1 & 0 & 0 & 0 & 1 & 1 \\
      0 & 0 & 1 & 0 & 0 & 0 & 0 & 0 & 0 & 0 & 1 & 0 \hspace*{2mm} |\hspace*{2mm} 1 & 0 & 0 & 1 & 0 & 0 & 0 & 1 & 1 & 0 & 0 & 1 \\
      0 & 0 & 0 & 1 & 0 & 0 & 0 & 0 & 0 & 0 & 0 & 1 \hspace*{2mm} |\hspace*{2mm} 1 & 1 & 0 & 0 & 1 & 0 & 0 & 0 & 1 & 1 & 0 & 0 \\
      1 & 0 & 0 & 0 & 1 & 0 & 0 & 0 & 0 & 0 & 0 & 0 \hspace*{2mm} |\hspace*{2mm} 0 & 1 & 1 & 0 & 0 & 1 & 1 & 0 & 0 & 1 & 0 & 0 \\
      0 & 1 & 0 & 0 & 0 & 1 & 0 & 0 & 0 & 0 & 0 & 0 \hspace*{2mm} |\hspace*{2mm} 0 & 0 & 1 & 1 & 0 & 0 & 1 & 1 & 0 & 0 & 1 & 0 \\
      0 & 0 & 1 & 0 & 0 & 0 & 1 & 0 & 0 & 0 & 0 & 0 \hspace*{2mm} |\hspace*{2mm} 1 & 0 & 0 & 1 & 1 & 0 & 0 & 1 & 0 & 0 & 0 & 1 \\
      0 & 0 & 0 & 1 & 0 & 0 & 0 & 1 & 0 & 0 & 0 & 0 \hspace*{2mm} |\hspace*{2mm} 1 & 1 & 0 & 0 & 1 & 1 & 0 & 0 & 1 & 0 & 0 & 0 \\
 \end{array}
  \right)
\end{equation}
\\
By removing the last four rows from the binary matrix $B$ we can
achieve $n-k=8$ independent generators. The distance $d$ of the
quantum code is given by the minimum weight of the bitwise OR
$\textbf{(a,b)}$ of all pairs satisfying the symplectic
orthogonality condition,
\\
\begin{equation}
B_1 \textbf{b} + B_2 \textbf{a}=0 \hspace*{3cm}
\end{equation}
\\
Let $\textbf{a}=(x_{01},...,x_{12})$ and
$\textbf{b}=(y_{01},...,y_{12})$. Then by using (4.3), we have
\\
\begin{equation}
\left\{
  \begin{array}{ll}
    x_{02} + x_{06} + x_{07} + x_{10} + x_{11} + y_{05} + y_{09} =0 \\
    x_{03} + x_{07} + x_{08} + x_{11} + x_{12} + y_{06} + y_{10} =0 \\
    x_{04} + x_{05} + x_{08} + x_{09} + x_{12} + y_{07} + y_{11} =0 \\
    x_{01} + x_{05} + x_{06} + x_{09} + x_{10} + y_{08} + y_{12} =0 \\
    x_{02} + x_{03} + x_{06} + x_{10} + x_{11} + y_{01} + y_{09} =0 \\
    x_{03} + x_{04} + x_{07} + x_{11} + x_{12} + y_{02} + y_{10} =0 \\
    x_{01} + x_{04} + x_{08} + x_{09} + x_{12} + y_{03} + y_{11} =0 \\
    x_{01} + x_{02} + x_{05} + x_{09} + x_{10} + y_{04} + y_{12} =0
\end{array}
\right.
 \hspace*{3cm}
\end{equation}
\\
By using (4.4) we can get the code distance $d$ equal to $3$. Since
the number of independent generators is $n-k=8$, therefore the
quantum stabilizer code is of length $12$, that encodes $k=4$
logical qubits, i.e., $[[ 12,4,3 ]]_{2}$ is constructed. This code
generated by the $n-k=8$ independent generators in table $9$.
\\
$$
\begin{tabular}{|c|c|}
  \hline
   Name & Operator\\
  \hline
  $g_1$ &I\hspace*{2mm}Z\hspace*{2mm}I\hspace*{2mm}I\hspace*{2mm}X\hspace*{2mm}Z\hspace*{2mm}Z\hspace*{2mm}I\hspace*{2mm}X\hspace*{2mm}Z\hspace*{2mm}Z\hspace*{2mm}I \\
  $g_2$ &I\hspace*{2mm}I\hspace*{2mm}Z\hspace*{2mm}I\hspace*{2mm}I\hspace*{2mm}X\hspace*{2mm}Z\hspace*{2mm}Z\hspace*{2mm}I\hspace*{2mm}X\hspace*{2mm}Z\hspace*{2mm}Z \\
  $g_3$ &I\hspace*{2mm}I\hspace*{2mm}I\hspace*{2mm}Z\hspace*{2mm}Z\hspace*{2mm}I\hspace*{2mm}X\hspace*{2mm}Z\hspace*{2mm}Z\hspace*{2mm}I\hspace*{2mm}X\hspace*{2mm}Z \\
  $g_4$ &Z\hspace*{2mm}I\hspace*{2mm}I\hspace*{2mm}I\hspace*{2mm}Z\hspace*{2mm}Z\hspace*{2mm}I\hspace*{2mm}X\hspace*{2mm}Z\hspace*{2mm}Z\hspace*{2mm}I\hspace*{2mm}X \\
  $g_5$ &X\hspace*{2mm}Z\hspace*{2mm}Z\hspace*{2mm}I\hspace*{2mm}I\hspace*{2mm}Z\hspace*{2mm}I\hspace*{2mm}I\hspace*{2mm}X\hspace*{2mm}Z\hspace*{2mm}Z\hspace*{2mm}I \\
  $g_6$ &I\hspace*{2mm}X\hspace*{2mm}Z\hspace*{2mm}Z\hspace*{2mm}I\hspace*{2mm}I\hspace*{2mm}Z\hspace*{2mm}I\hspace*{2mm}I\hspace*{2mm}X\hspace*{2mm}Z\hspace*{2mm}Z \\
  $g_7$ &Z\hspace*{2mm}I\hspace*{2mm}X\hspace*{2mm}Z\hspace*{2mm}I\hspace*{2mm}I\hspace*{2mm}I\hspace*{2mm}Z\hspace*{2mm}Z\hspace*{2mm}I\hspace*{2mm}X\hspace*{2mm}Z \\
  $g_8$ &Z\hspace*{2mm}Z\hspace*{2mm}I\hspace*{2mm}X\hspace*{2mm}Z\hspace*{2mm}I\hspace*{2mm}I\hspace*{2mm}I\hspace*{2mm}Z\hspace*{2mm}Z\hspace*{2mm}I\hspace*{2mm}X \\
  \hline
\end{tabular}
$$
\begin{table}[htb]
\caption{\small{Stabilizer generators for the $[[12,4,3]]_{2}$
code.}} \label{table:1}
\newcommand{\m}{\hphantom{$-$}}
\newcommand{\cc}[1]{\multicolumn{1}{c}{#1}}
\renewcommand{\tabcolsep}{0pc} 
\renewcommand{\arraystretch}{1.} 
\end{table}
\\
Applying (2.27), (2.31), (2.35), (2.38) and (2.39) we can obtain
quantum stabilizer codes from $U_{6n}$, $T_{4n}$, $V_{8n}$ and
dihedral $D_{2n}$ groups. A list of quantum stabilizer codes is
given in table $10$.
\\
\\
\textbf{Remark.} Table $10$ is a list of quantum stabilizer codes
from $U_{6n}$, $T_{4n}$, $V_{8n}$ and dihedral $D_{2n}$ groups. The
first column shows non-Abelian groups. The second column shows $B_1$
and $B_2$ in terms of $A_i$, $i=0,1,...,m$. where $m$ denotes the
number of conjugacy classes of the group $G$. The third column shows
the value of the length of quantum stabilizer code. The fourth
column shows the value of $n-k$. The fifth column shows a list of
the quantum stabilizer codes.
\\
\newpage
$$
\begin{tabular}{|c|p{10.5cm}|c|c|l|}
  \hline
\hline
  Group  & $B_i(i=1,2)$ & $n$ & $n-k$ & $[[ n,k,d ]]_{2}$\\
  \hline
 $U_{12}$ & $B_1=A_1+A_2+A_4$,\hspace*{2mm}$B_2=A_3$ & $12$ & $8$ & $[[ 12,4,3 ]]_{2}$ \\
 $U_{12}$ & $B_1=A_1+A_2+A_5$,\hspace*{2mm}$B_2=A_0+A_4$ & $12$ & $8$ & $[[ 12,4,3 ]]_{2}$ \\
 $U_{12}$ & $B_1=A_2$,\hspace*{2mm}$B_2=A_3+A_5$ & $12$ & $8$ & $[[ 12,4,3 ]]_{2}$ \\
 $U_{12}$ & $B_1=A_1+A_2+A_5$,\hspace*{2mm}$B_2=A_0+A_4$ & $12$ & $11$ & $[[ 12,1,4 ]]_{2}$ \\
 $U_{18}$ & $B_1=A_1+A_2+A_3+A_7+A_8$,
\hspace*{2mm}$B_2=A_0+A_1+A_2+A_4+A_5$ & $18$ & $12$ & $[[ 18,6,3]]_{2}$ \\
$U_{18}$ & $B_1=A_1+A_2+A_3+A_7$,
\hspace*{2mm}$B_2=A_0+A_1+A_2+A_4$ & $18$ & $13$ & $[[ 18,5,3]]_{2}$ \\
$U_{18}$ & $B_1=A_1+A_2+A_3+A_7$,
\hspace*{2mm}$B_2=A_0+A_1+A_2+A_4$ & $18$ & $16$ & $[[ 18,2,4]]_{2}$ \\
 $U_{24}$ & $B_1=A_0+A_1+A_2+A_3+A_4+A_8+A_{10}$,\hspace*{2mm}$B_2=A_0+A_3+A_5+A_6+A_{11}$ & $24$ & $12$ & $[[ 24,12,3 ]]_{2}$ \\
 $U_{24}$ & $B_1=A_0+A_1+A_2+A_3+A_4+A_8+A_{10}$,\hspace*{2mm}$B_2=A_0+A_3+A_5+A_6+A_{11}$ & $24$ & $16$ & $[[ 24,8,5 ]]_{2}$ \\
 $T_{12}$ & $B_1=A_2+A_4$,\hspace*{2mm}$B_2=A_0+A_5$ & $12$ & $9$ & $[[ 12,3,3 ]]_{2}$ \\
 $T_{12}$ & $B_1=A_0+A_4$,
\hspace*{2mm}$B_2=A_1+A_2+A_5$ & $12$ & $10$ & $[[ 12,2,3 ]]_{2}$ \\
 $T_{16}$ & $B_1=A_0+A_1+A_2+A_6$,\hspace*{2mm}$B_2=A_0+A_2+A_3$ & $16$ & $14$ & $[[ 16,2,3 ]]_{2}$ \\
 $V_{24}$ & $B_1=A_0+A_3+A_6+A_7$,\hspace*{2mm}$B_2=A_0+A_2+A_4$ & $24$ & $20$ & $[[ 24,4,3 ]]_{2}$ \\
 $D_{12}$ & $B_1=A_3+A_5$,\hspace*{2mm}$B_2=A_2+A_3+A_5$ & $12$ & $10$ & $[[ 12,2,3 ]]_{2}$ \\
\hline \hline
\end{tabular}
$$
\begin{table}[htb]
\caption{\small{Quantum stabilizer codes $[[ n,k,d ]]_{2}$.}}
\label{table:2}
\newcommand{\m}{\hphantom{$-$}}
\newcommand{\cc}[1]{\multicolumn{1}{c}{#1}}
\renewcommand{\tabcolsep}{0pc} 
\renewcommand{\arraystretch}{1.} 
\end{table}
\\
\\
\section{\hspace*{-.5cm}\ Conclusion}
We have developed a new method of constructing binary quantum
stabilizer codes from Abelian and non-Abelian groups association
schemes. Using this method, we have constructed good binary quantum
stabilizer codes of distances $3$, $4$, $5$, and $7$ up to $40$.
Furthermore, binary quantum stabilizer codes of a large length $n$
with high distance can be constructed. We can see from tables 4 and
5 that the Abelian association schemes procedure for the
construction of the binary quantum stabilizer codes is superior to
non-Abelian group association schemes. Although we focused
specifically on Abelian and non-Abelian groups association schemes,
we expect that the introduced method might then be applied to other
association schemes such as association scheme defined over the
coset space $G/H$, where $H$ is a normal subgroup of finite group
$G$ with prime index., strongly regular graphs, distance regular
graphs, etc. These association schemes are under investigation.
\\

\end{document}